\documentclass[floatfix,twocolumn,showpacs,prc]{revtex4-1}

\usepackage{lipsum}
\usepackage[utf8]{inputenc}
\usepackage[english]{babel}
\usepackage{tikz}
\usepackage{paralist}
\usepackage{subcaption} \usepackage{graphicx}
\usepackage{amsmath} \usepackage{amssymb}
\usepackage{booktabs}
\usepackage{xcolor}
\usepackage{todonotes}
\usepackage{float}
\usepackage[inline]{enumitem}
\graphicspath{{./fig/}}

\begin{document}

\title[Fragment Formation in Neutron Rich Matter]{Dynamics of Fragment Formation in Neutron Rich Matter}

\author{P. N. Alcain and C. O. Dorso}

\affiliation{Departamento de Física, FCEyN, UBA and IFIBA, Conicet, Pabellón 1, Ciudad Universitaria, 1428 Buenos Aires, Argentina}
\affiliation{IFIBA-CONICET}

\date{\today}

\begin{abstract}
  \begin{description}
  \item[Background] Neutron stars are astronomical systems with nucleons submitted to extreme conditions.
    Due to the longer range coulomb repulsion between protons, the system has structural inhomogeneities.
    Several interactions tailored to reproduce nuclear matter plus screened Coulomb term reproduce these inhomogeneities known as \emph{nuclear pasta}.
    These structural inhomogeneities, located in the crust of neutron stars, can also arise in expanding systems depending on the thermodynamic conditions (temperature, proton fraction, \ldots) and the expansion velocity.

\item[Purpose] We aim to find the dynamics of the fragments formation for expanding systems simulated according to the little big bang model.
  This expansion resembles the evolution of neutron stars merger.
  
\item[Method] We study the dynamics of the nucleons with semiclassical molecular dynamics models.
  Starting with an equilibrium configuration, we expand the system homogeneously until we arrive to an asymptotic configuration (i.\ e.\ very low final densities).
  We study, with four different cluster recognition algorithms, the fragment distribution throughout this expansion and the dynamics of the cluster formation.
  
\item[Results] Studying the topology of the equilibrium states, before the expansion, we reproduced the known pasta phases plus a novel phase we called \emph{pregnocchi}, consisting of proton aggregates embedded in a \emph{neutron sea}.
  We have identified different fragmentation regimes, depending on the initial temperature and fragment velocity.
  In particular, for the already mentioned \emph{pregnocchi}, a neutron cloud surrounds the clusters during the early stages of the expansion, resulting in systems that give rise to configurations compatibles with the emergence of \emph{r-proccess}.
  
\item[Conclusions] These calculations pave the way to a comparision between Earth experiments and neutron star studies.
  \end{description}
  
\end{abstract}

\maketitle

\section{Introduction}\label{sc:intro}
Neutron rich matter is present in several astronomical objects in the universe, for example: neutron stars, proto-neutron stars and core-collapse supernovae.
The supernova explosion of a massive star, combined with gravitational collapse, compresses the core up to densities of atomic nuclei.
This gives rise to a system known as \emph{proto-neutron star}, which eventually ends up in a neutron star.

The neutron rich environment also gives rise to the possibility of a rapid neutron capture, \emph{r}-process, that consists of the rapid capture of neutrons.
The \emph{r}-process is fundamental to understand the abundancy of heavy elements, and several places have been candidates for it to happen.
Supernovae have been prime candidates for long, but recent observations and models hint that also neutron star mergers can yield \emph{r}-process~\cite{eichler_nucleosynthesis_1989, freiburghaus_r-process_1999, tanvir_/`kilonova/_2013}.
The compression of neutron star matter as a possible source for \emph{r}-process nuclei was first discussed in Ref.~\cite{lattimer_black-hole-neutron-star_1974}.
According to hydrodynamic models~\cite{goriely_r-process_2011}, these have typically velocity gradients of $\dot{\eta} = 10^{-21}\,\text{c/fm} < \dot{\eta} < 4\cdot 10^{-20}\,\text{c/fm}$.

The original works of Ravenhall \emph{et al.}~\cite{ravenhall_structure_1983} and Hashimoto \emph{et al.}~\cite{hashimoto_shape_1984} used a compressible liquid drop model to study neutron rich matter, and have shown that the now known as the \emph{pasta phases} --\emph{lasagna}, \emph{spaghetti} and \emph{gnocchi}-- are solutions to the ground state of neutron star matter.
The study of neutron rich matter has since been approached with different models, which show that \emph{nuclear pasta} arises due to the interplay between nuclear and Coulomb forces in an infinite medium.
We classify the different approaches in two large groups: mean field and microscopic.

Mean field works include the Liquid Drop Model, by Lattimer \emph{et al.}~\cite{page_minimal_2004}, Thomas-Fermi, by Williams and Koonin~\cite{williams_sub-saturation_1985}, among others~\cite{oyamatsu_nuclear_1993, lorenz_neutron_1993, cheng_properties_1997, watanabe_thermodynamic_2000, watanabe_electron_2003, nakazato_gyroid_2009}.
Microscopic models include Quantum Molecular Dynamics, used by Maruyama \emph{et al.}~\cite{maruyama_quantum_1998, kido_md_2000} and by Watanabe~\emph{et al.}\cite{watanabe_structure_2003}, Simple Semiclassical Potential, by Horowitz~\emph{et al.}~\cite{horowitz_nonuniform_2004} and Classical Molecular Dynamics, used in our previous works~\cite{dorso_topological_2012}.

In some recent studies, phases different from the typical \emph{nuclear pasta} were found.
The work by Nakazato \emph{et al.}~\cite{nakazato_gyroid_2009}, inspired by polymer systems, found also gyroid and double-diamond structures, with a compressible liquid drop model.
Dorso \emph{et al.}~\cite{dorso_topological_2012} obtained pasta phases different from those already mentioned with molecular dynamics, studying mostly their characterization at very low temperatures.
In our previous work~\cite{alcain_beyond_2014} we have shown that these new pasta phases had an opacity peak (i.\ e., a local maximum in the opacity) in the characteristic wavelength of the Urca neutrinos for symmetrical neutron star matter.

Among the advantages of classical and semiclassical models are the accessibility to position and momentum of all particles at all times, which allows the calculation of correlations of all orders.
Moreover, no specific structure is hardcoded in the model, as it happens with most mean field models.
This enables the study of the structure of the nuclear medium from a particle-wise point of view.
Many models exist with this goal, including quantum molecular dynamics~\cite{maruyama_quantum_1998}, simple-semiclassical potential~\cite{horowitz_nonuniform_2004} and classical molecular dynamics~\cite{lenk_accuracy_1990}.
In these models the Pauli repulsion between nucleons of equal isospin is hard-coded in the interaction.
On the other hand, a specific Pauli potential developed in~\cite{dorso_classical_1987} was used in the QCNM~\cite{dorso_classical_1988} and later in Ref.~\cite{hartnack_quantum_1989}.

The relative inaccessibility to these astronomical objects means a restriction in the observables available.
One of them, studied extensively in the recent years, is the neutrino opacity and the mean free path~\cite{horowitz_neutrino-pasta_2004, horowitz_nonuniform_2004, alcain_neutrino_2017}.
In this work, we study another possible observable from the neutron rich matter: the result of the fragmentation of neutron rich matter, related to the already mentione \emph{r-process}.
Multifragmentation in nuclear systems has been studied before~\cite{bonasera_critical_2000, chikazumi_quantum_2001}, but mostly with nuclear matter (without Coulomb interaction).
In a recent work by Caplan et al~\cite{caplan_pasta_2015}, expanding neutron star matter has been studied as possible explanations for nucleosynthesis in neutron star mergers.

In Section~\ref{sc:model} we introduce the model used along this work, that includes the potential parametrization (\ref{ssc:cmd}) and the Coulomb interaction (\ref{ssc:coulomb}).
Section~\ref{sc:cluster} describes the different cluster recognition algorithms used in this work, and Section~\ref{sc:expansion} explains  how we simulate the expansion of the system.
Finally, we draw conclusions in Section~\ref{sc:conc}.
In the appendix we perform a detailed analysis on the stability of one of the cluster recognition algorithms.

\section{The Model}\label{sc:model}

\subsection{Classical Molecular Dynamics}\label{ssc:cmd}
In this work, we study fragmentation of Neutron Star Matter under pasta-like conditions with model similar to the classical molecular dynamics model CMD.\@
CMD has been used in several heavy-ion reaction studies to: help understand experimental data~\cite{chernomoretz_quasiclassical_2002}; identify phase-transition signals and other critical phenomena~\cite{lopez_lectures_2000, barranon_searching_2001, dorso_selection_2001, barranon_critical_2003, barranon_time_2007}; and explore the caloric curve~\cite{barranon_entropy_2004} and isoscaling~\cite{dorso_dynamical_2006, dorso_isoscaling_2011}.
CMD uses two two-body potentials to describe the interaction of nucleons, which are a combination of Yukawa potentials:
\begin{align*}
  V^{\text{CMD}}_{np}(r) &=v_{r}\exp(-\mu_{r}r)/{r}-v_{a}\exp(-\mu_{a}r)/{r}\\
  V^{\text{CMD}}_{nn}(r) &=v_{0}\exp(-\mu_{0}r)/{r}
\end{align*}
where $V_{np}$ is the potential between a neutron and a proton, and $V_{nn}$ is the repulsive interaction between either $nn$ or $pp$.
The cutoff radius is $r_c=5.4\,\text{fm}$ and for $r>r_c$ both potentials are set to zero.
The Yukawa parameters $\mu_r$, $\mu_a$ and $\mu_0$ were determined to yield an equilibrium density of $\rho_0=0.16 \,\text{fm}^{-3}$, a binding energy $E(\rho_0)=16 \,\text{MeV/nucleon}$ and a compressibility of $250\,\text{MeV}$.

\begin{figure}[h]
  \centering
  \includegraphics[width=0.8\columnwidth]{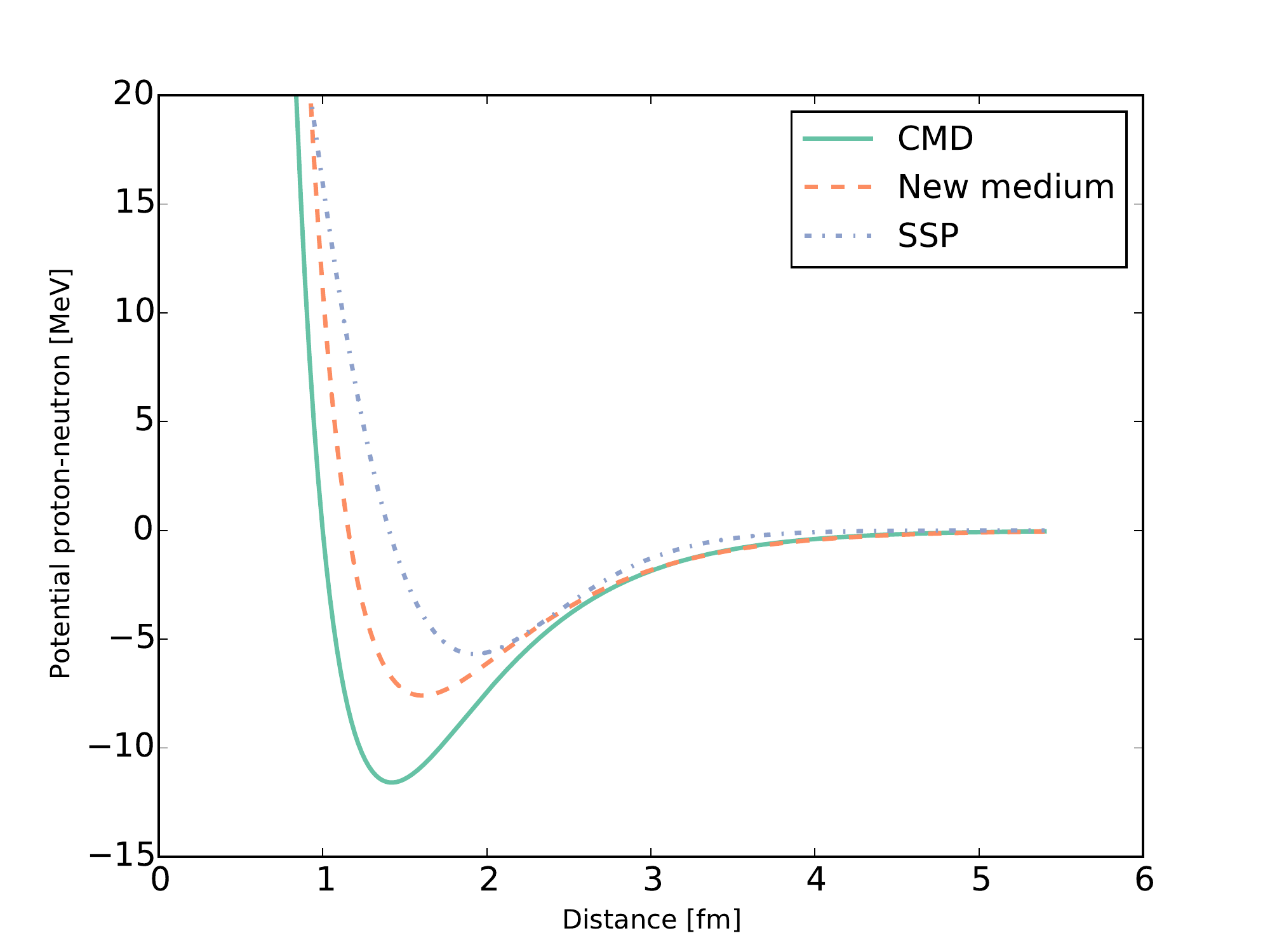}
  \caption{Potential energy of the proton-neutron interaction of different models: SSP, CMD and New Medium.}
\label{fig:vnp}
\end{figure}

Based on this model, we developed a new set of parameters that yield the same values for $\rho_0$, $E(\rho_0)$ and compressibility, which we called \emph{New Medium}.
We show in figure~\ref{fig:vnp} an example that compares the proton-neutron potential for the different models and the developed for this work: SSP, CMD and New Medium.

To simulate an infinite medium, we used this potential with $N = 5500$ particles under periodic boundary conditions, with different proton fractions (i.\ e.\ with $x = Z/A = 0.2 < x < 0.4$) in cubical boxes with sizes adjusted to have densities $\rho=0.05 \,\text{fm}^{-3}$ and $\rho = 0.08\,\text{fm}^{-3}$.
These simulations have been done with LAMMPS~\cite{plimpton_fast_1995}, using its GPU package~\cite{brown_implementing_2012}.

\subsubsection{Ground State Nuclei}
\begin{figure}[h]
  \centering
  \includegraphics[width=0.8\columnwidth]{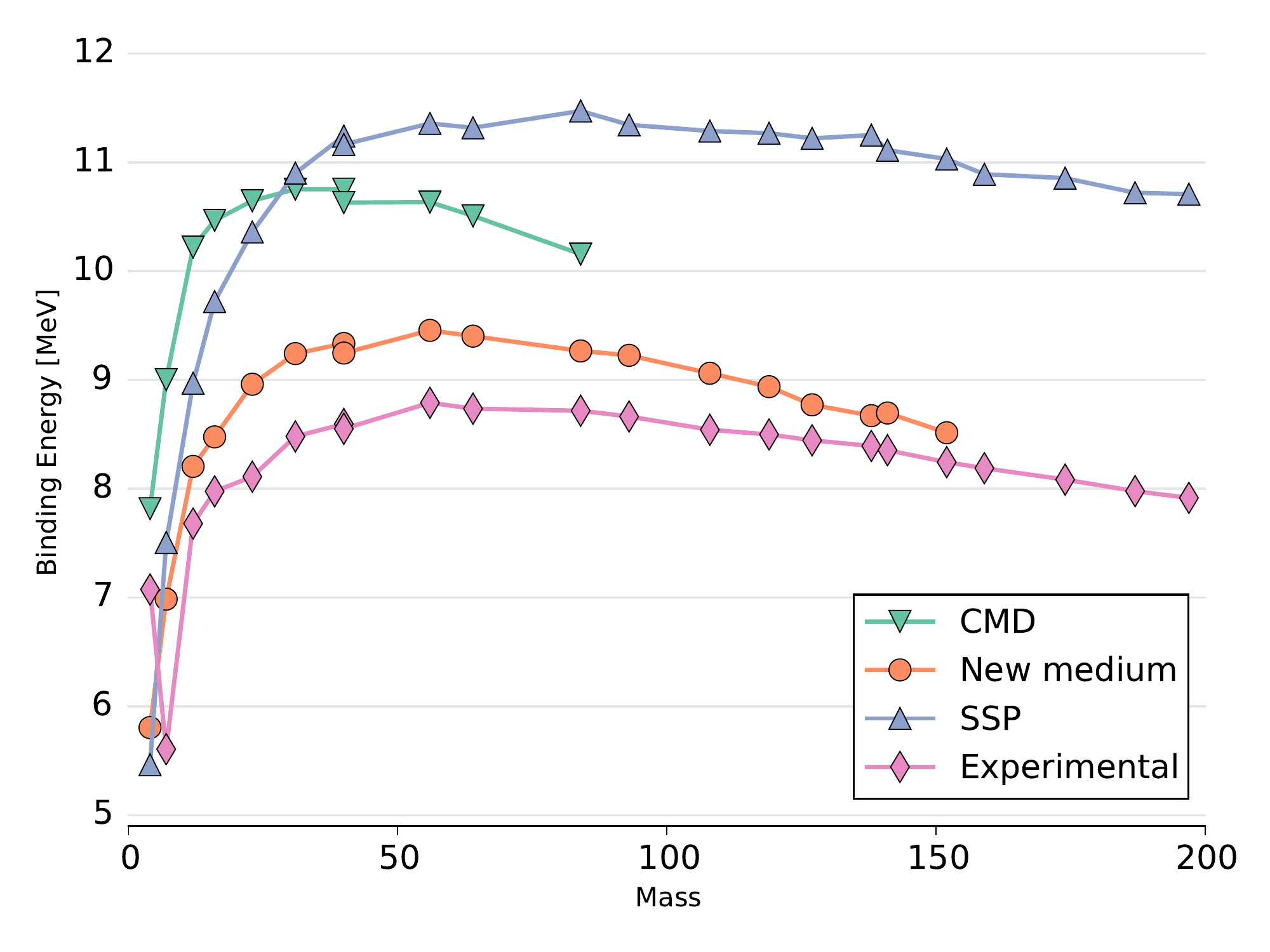}
  \caption{Binding energies of ground-state nuclei obtained with CMD, SSP and New Medium model. See that the New Medium model yields results much closer to the experimental ones.}
\label{fig:binding}
\end{figure}

Although the $T=0$ state of this classical nuclear matter at normal densities is a simple cubic solid, nuclear systems can be mimicked by adding enough kinetic energy to the nucleons.
To study nuclei, for instance, liquid-like spherical drops with the right number of protons and neutrons are constructed confined in a steep spherical potential and then brought to the ground state by cooling them slowly from a rather high temperature until they reach a self-contained state.
Removing the confining potential, the system is further cooled down until a reasonable binding energy is attained.
The remaining kinetic energy of the nucleons helps to resemble the Fermi motion.
To compare the different microscopic models used throughout the literature with the \emph{New Medium}, we show in figure~\ref{fig:binding} the binding energies of ground-state nuclei obtained with CMD, SSP and New Medium.

\subsection{Coulomb interaction in the model}\label{ssc:coulomb}

Since a neutralizing electron gas embeds the nucleons in the neutron star crust, the Coulomb forces among protons are screened.
We model this screening effect with the Thomas-Fermi approximation, used with various nuclear models~\cite{maruyama_quantum_1998, dorso_topological_2012, horowitz_neutrino-pasta_2004}.
According to this approximation, protons interact via a Yukawa-like potential, with a screening length $\lambda$:
\begin{equation}
 V_{TF}(r) = q^2\frac{e^{-r/\lambda}}{r}
\end{equation}

Theoretical estimates for the screening length $\lambda$ are $\lambda\sim100\,\text{fm}$~\cite{fetter_quantum_2003}, but we set the screening length to $\lambda=20\,\text{fm}$.
This choice was based on previous studies~\cite{alcain_effect_2014}, where we have shown that this value is enough to adequately reproduce the expected length scale of density fluctuations for this model, while larger screening lengths would be a computational difficulty.
We analyze the opacity to neutrinos of the structures for different proton fractions and densities.

\section{Cluster recognition}\label{sc:cluster}
In typical configurations we have not only the structure known as nuclear pasta, but also a nucleon gas that surrounds the nuclear pasta.
In order to properly characterize the pasta phases, we must know which particles belong to the pasta phases and which belong to this gas.
To do so, we have to find the clusters that are formed along the simulation.

One of the algorithms to identify cluster formation is Minimum Spanning Tree (MST).
In MST algorithm, two particles belong to the same cluster $\{C^{\text{MST}}_n\}$ if the relative distance of the particles is less than a cutoff distance $r_{cut}$:
\begin{equation*}
  i \in C^{\text{MST}}_n \Leftrightarrow \exists j \in C_n \mid
  r_{ij} < r_{cut}
\end{equation*}

Based on MST algorithm, and taking into account that typically the neutron rich matter structure is set by the proton backbone, we developed MSTpC:\@ an algorithm that calculates the MST cluster of protons alone and finds the cloud of neutrons that lay within $r_c$ of each proton cluster.
The MST cluster definition works correctly for systems with no kinetic energy, and it is based in the attractive tail of the nuclear interaction.
However, if the particles have non-zero relative momenta, we can have a situation of two particles that are closer than the cutoff radius, but with a large relative kinetic energy.

The problem of fragment recognition in nuclear physics has had a strong development in the last years, specially related to the analysis of numerical simulations of intermediate energy heavy ion collisions.
In this case, it is quite clear that the asymptotic state of the system is a very dilute set of fragments with a collective expansion mode and composed by cold fragments.
In the asymptotic state, fragments will be far away from each other, and as such, application of the above mentioned MST algorithm yields an accurate description of the fragmentation.
However, if one is interested in the analysis of the time evolution of the fragment structure, it is clear that the MST will not provide information because during the expansion nearby particles may have very different momenta, which are not considered in MST definition.
Therefore, two particles can be very close to each other --within range of the attractive potential-- but with high relative momentum being recognized as a bound pair according to MST.\@
This unwanted behavior can be partially solved using the MSTE algorithm, in which two particles belong to the same cluster $\{C^{\text{MSTE}}_n\}$ if they are energy bound:
\begin{equation*}
  i \in C^{\text{MSTE}}_n \Leftrightarrow \exists j \in C^{\text{MSTE}}_n :
  V_{ij}+ K_{ij} \le 0
\end{equation*}
This is an approximate solution to our problem, but it shows signs of instability even in some simple cases (see appendix).

One of the most sophisticated methods to find the energetically bound clusters is the Early Cluster Recognition Algorithm (ECRA)~\cite{dorso_early_1993}.
In this algorithm, the particles are partitioned in different disjoint clusters $C^{\text{ECRA}}_n$, with the total energy in each cluster:
\begin{equation*}
  \epsilon_n = \sum_{i \in C_n} K^{CM}_i +  \sum_{i,j \in C_n} V_{ij}
\end{equation*}
where $K^{CM}_i$ is the kinetic energy relative to the center of mass of the cluster.
The set of clusters $\{C^{\text{ECRA}}_n\}$ then is the one that minimizes the sum of all the cluster energies $E_{\text{partition}} = \sum_n \epsilon_n$.

As mentioned above, expanding systems have the property that the asymptotic state is easy to calculate, so the efficiency of other recognition algorithms apart from MST can be easily tested.
Moreover, the quality of the recognition algorithm can be weighted considering how early in the evolution of the expanding system it is able to identify the clusters corresponding to the asymptotic state.
In this sense, the ECRA algorithm has shown that it is able to recognize fragments very early in the evolution, providing a new view of the dynamics of the fragment formation i.\ e., fragments are early formed in exploding systems.
ECRA algorithm can be easily used for small systems~\cite{dorso_fluctuation_1994}, but being a combinatorial optimization, it cannot be used in large systems.

To find approximate solutions, the originally proposed method is similar to simulated annealing~\cite{dorso_early_1993}.
Another other choice was developed by Puente~\cite{puente_efficient_1999}, and it introduces a Binary Fusion Model.
In this model, the initial configuration is with all clusters being monomers (one particle per cluster).
With this starting point, $E_{\text{partition}}^0 = 0$ the steps that follow are:
\begin{enumerate}
\item Explore all potential mergers of two clusters and bookkeep the resulting $E_{\text{partition}}^{i+1}$ from each potential merger.
\item Pick the merger that results in the lowest $E_{\text{partition}}^{i+1}$.
\item If $E_{\text{partition}}^{i+1} < E_{\text{partition}}^i$, perform the merge and go back to step 1; otherwise, stop iteration.
\end{enumerate}

All of these algorithms for cluster recognition should give the same results for the asymptotic state.

\subsection{Infinite Clusters}

\begin{figure}  \centering
  \includegraphics[width=0.8\columnwidth]{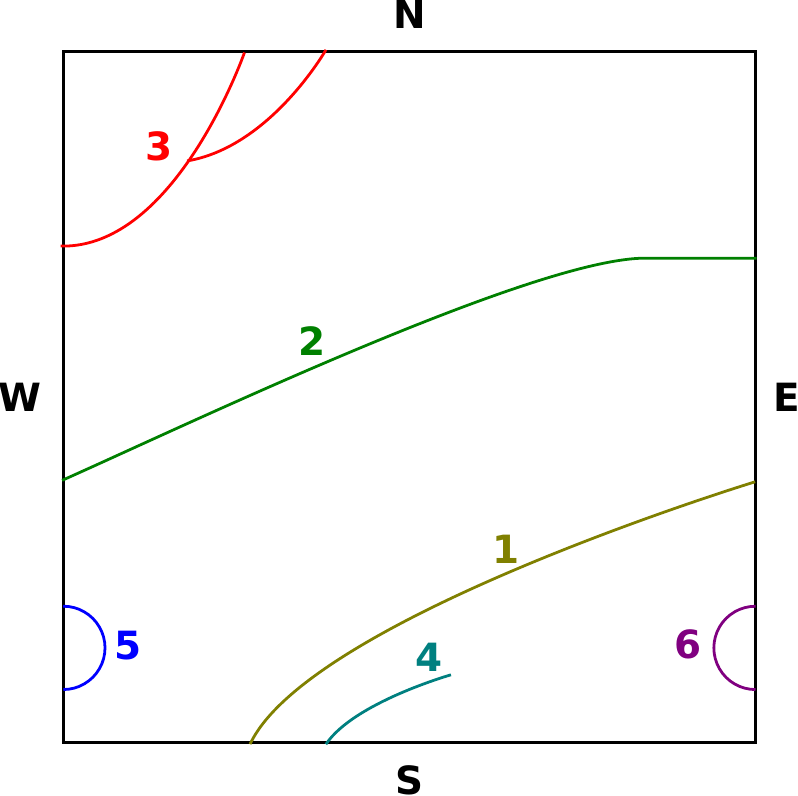}
  \caption{(Color online) Schematical representation of 2D clusters, recognized only in the cell and not through the periodic walls, labeled as N, S, W, E.
    The clusters inside the cell are labeled from 1 to 6.}
\label{fig:scheme_clusters}
\end{figure}

\begin{figure}  \centering
  \includegraphics[width=0.45\columnwidth]{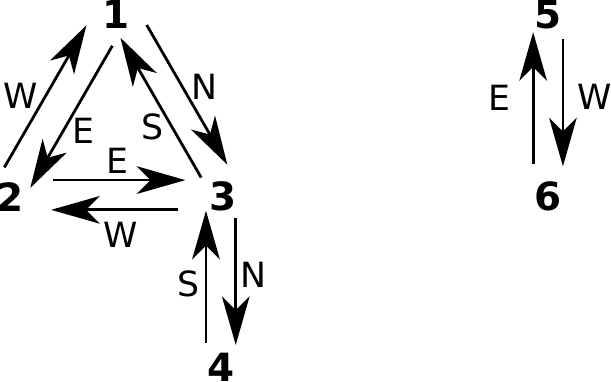}
  \caption{Graph of the clusters with connections labeled by the wall of the boundary they connect through.
    The graph can be divided in  2 subgraphs that don't connect: 1--2--3--4 and 5--6.
    Each of these subgraphs is as cluster when periodic boundary conditions are considered.}
\label{fig:graph_clusters}
\end{figure}

In Ref.~\cite{alcain_fragmentation_2016} we developed an algorithm for the recognition of infinite clusters across the boundaries.
We explain here in detail the implementation for MST clusters in 2D, being the MSTE and 3D extension straightforward.
In figure~\ref{fig:scheme_clusters} we see a schematical representation of 2D clusters recognized in a periodic cell, labeled from 1 to 6 (note that these clusters don't connect yet through the periodic walls).

In order to find the connections of these clusters through the boundaries, we draw a labeled graph of the clusters, where we connect clusters depending on whether they connect or not through a wall and label such connection with the wall label.
For example, we begin with cluster 1.
It connects with cluster 2 going out through the E wall, therefore we add a $1\rightarrow2$ connection labeled as E.
Symmetrically, we add a $2\rightarrow1$ connection labeled as W.
Now we go for the pair 1--3.
It connects going out through the S wall, so we add $1\rightarrow3$ labeled as S and $3\rightarrow1$ labeled as N.
Cluster 1 does not connect with 4, 5, or 6, therefore those are the only connections we have.
Once we've done that, we get the graph of figure~\ref{fig:graph_clusters}.

We now wonder whether these subgraphs represent an infinite cluster or not.
In order to have an infinite clusters, we need to have a loop (the opposite is not true: having a loop is not enough to have an infinite cluster, as we can see in subragph 5--6), so we first identify loops and mark them as candidates for infinite clusters.
Every connection adds to a loop (since the graph connections are back and forth), but we know from inspecting the figure~\ref{fig:graph_clusters} that the cluster 1--2--3 is infinite.
Finding out what makes, in the graph, the cluster 1--2--3 infinite is key to identify infinite clusters.
And the key feature of cluster 1--2--3 is that its loop 1--2--3--1 can be transversed through the walls E--E--S, while loops like 5--6 can be transversed only through E--W.
Now, in order for the cluster to be infinite, we need it to extend infinitely in (at least) one direction.
So once we have the list of walls of the loop, we create a magnitude $I$ associated to each loop that is created as follows: beginning with $I = 0$, we add a value $M_i$ if there is (at least one) $i$ wall.
The values are: $M_E = 1$, $M_W = -1$, $M_N = 2$, $M_S = -2$.
If $I$ is nonzero, then the loop is infinite.
For example, for the loop E--E--S, we have E and S walls, so $I = M_E + M_S = 3$ and the loop is infinite.
For the loop E--W, $I = M_E + M_W = 0$, and the loop is finite.

\section{Expansion}\label{sc:expansion}
In order to expand the neutron rich matter that simulates an infinite system with periodic boundary conditions, we follow the \emph{microscopic big bang} method, as explained by Dorso and Strachan in Ref.~\cite{dorso_onset_1996}.
It consists of an expansion of the simulation box at a constant isotropic rate:
\begin{equation}
  L(t) = L_0\,(1+\dot{\eta}\,t)
\end{equation}
where $L$ is the length of the simulation box in every direction and $L_0$ is the initial length.
With only this box resizing, the system would expand dynamically.
To simulate an expansion, we need to also give the particles an extra radial velocity that maches that of the box in the edges of the simulation:
\begin{equation}
  \mathbf{v} = \mathbf{v_0} + \dot{\eta}\,\mathbf{r_0}
\end{equation}

Since we are working with periodic boundary conditions, when a particle crosses a boundary, we must take into account the original expansion, so we do not only change the particle position but also the
velocity.
For example, if the particle crosses the left-hand boundary of the periodic box, the velocity of the image particle $v_i^\dagger$ on the right-hand must be modified $v_i^\dagger = v_i + L_0\,\dot{\eta}$.
This prescription for an expansion is mathematically equivalent to Hubble's law in astrophysics~\cite{chikazumi_quantum_2001}.

\section{Results}
\subsection{Configuration dependence with the potential}

\begin{figure*}
  \begin{subfigure}{.3\linewidth}
    \includegraphics[width=\columnwidth]{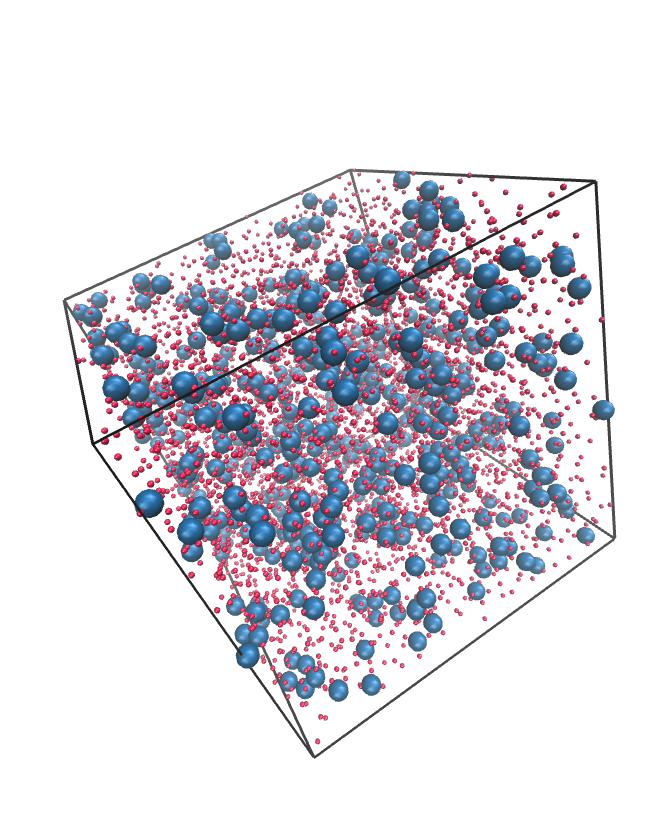}
    \caption{CMD medium}
  \end{subfigure}
  \begin{subfigure}{.3\linewidth}
    \includegraphics[width=\columnwidth]{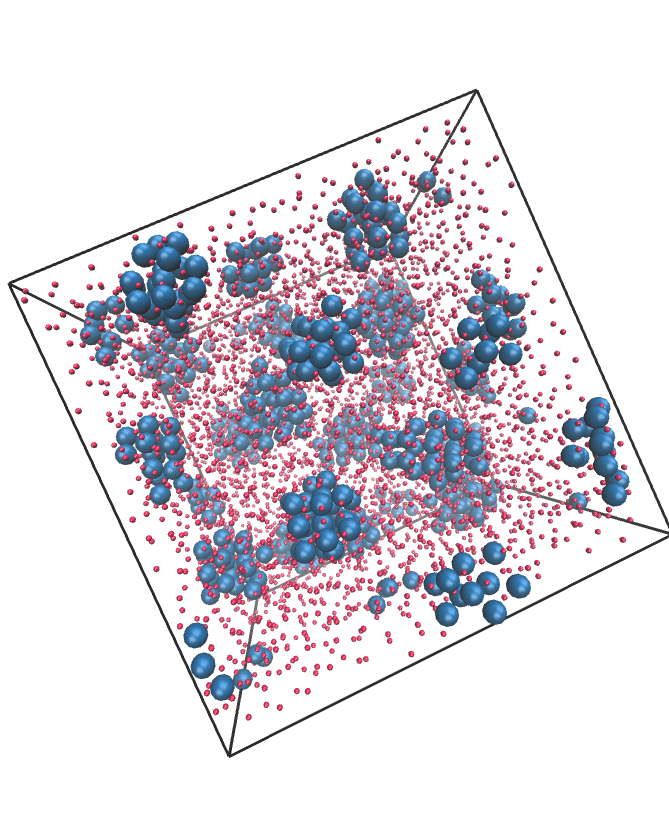}
    \caption{New medium}
  \end{subfigure}
  \begin{subfigure}{.3\linewidth}
    \includegraphics[width=\columnwidth]{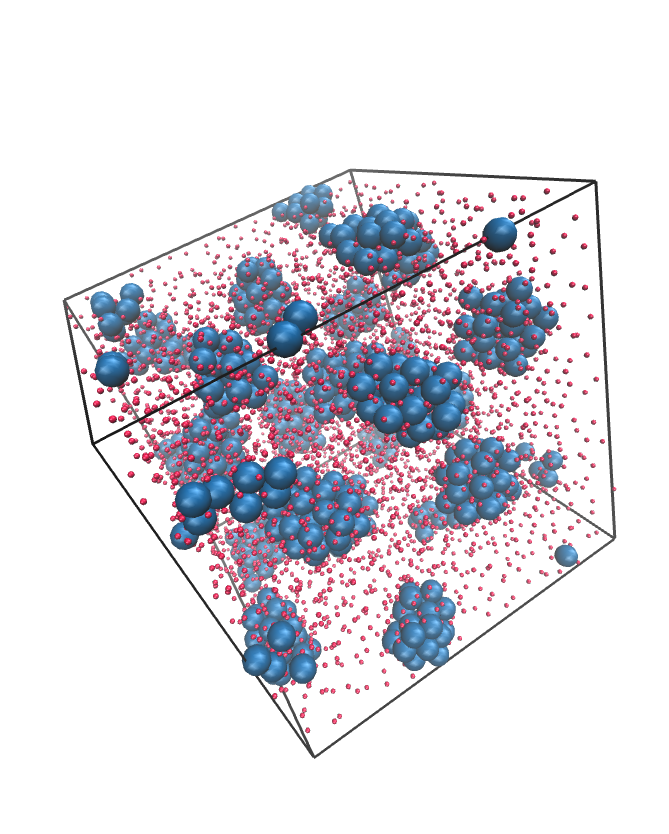}
    \caption{SSP}
  \end{subfigure}
  \caption{Snapshots of configurations for different parametrizations of the nuclear interaction, all with the same thermodynamic conditions: $x = 0.1$, $\rho = 0.05\,\text{fm}^{-3}$ and $T = 0.1\,\text{MeV}$.
    The qualitative differences between CMD medium potential and the other two parametrizations (New Medium and SSP) are evident.
    We call the structures shown in New Medium and SSP \emph{pregnocchi}.
    Please notice that heneutrons are repressented by points to avoid hindering the visualization of the proton structure.}
\label{fig:x01_potentials}
\end{figure*}

\begin{figure*}
  \begin{subfigure}{.3\linewidth}
    \includegraphics[width=\columnwidth]{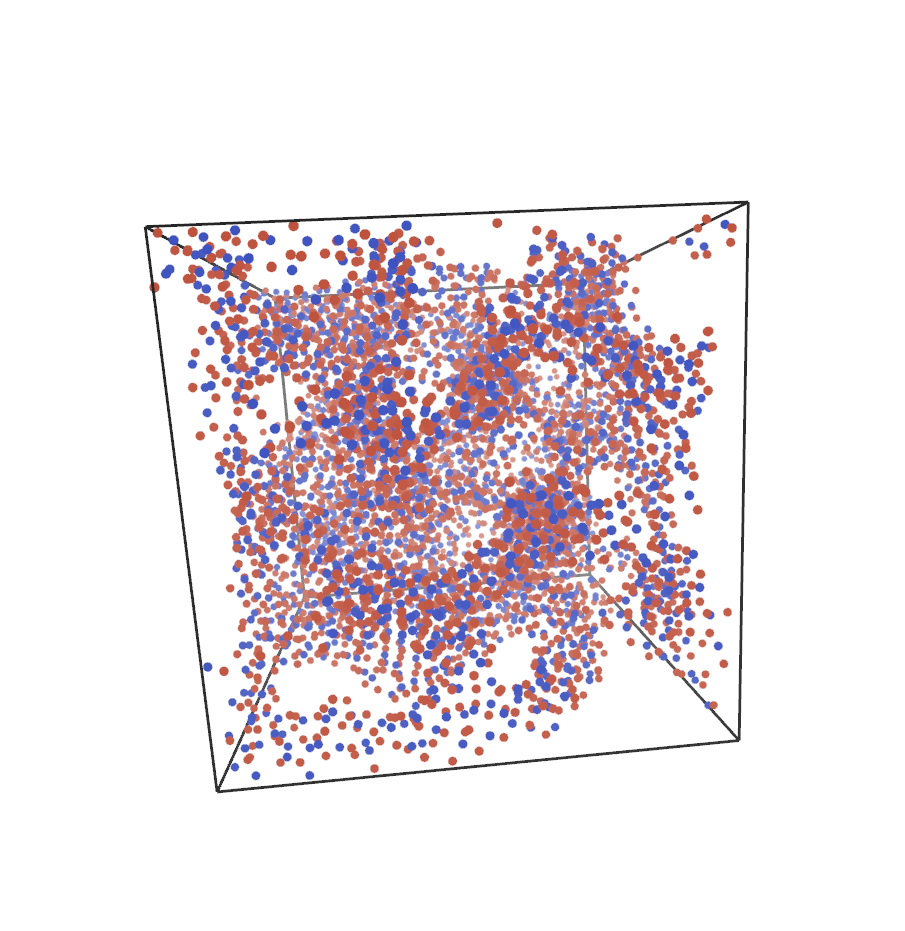}
    \caption{Medium}
  \end{subfigure}
  \begin{subfigure}{.3\linewidth}
    \includegraphics[width=\columnwidth]{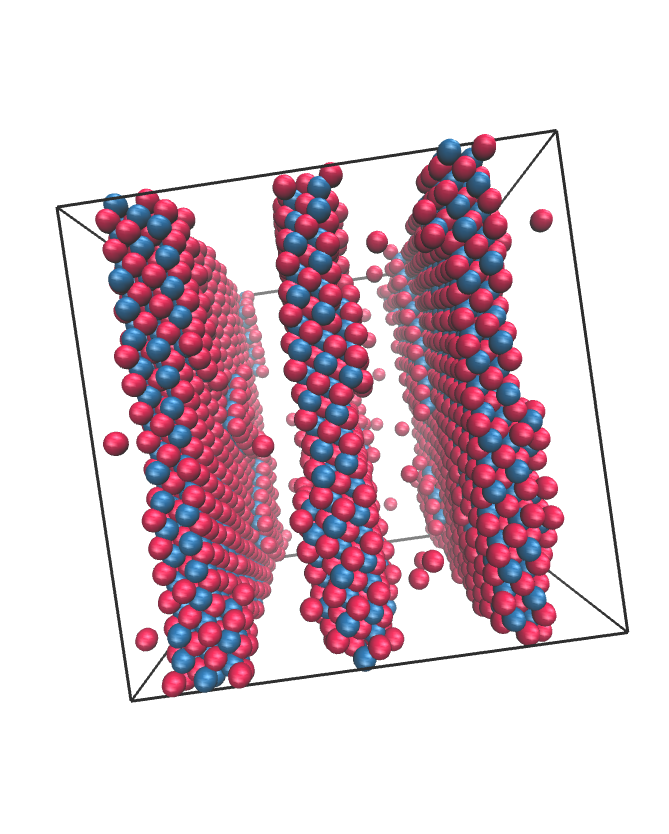}
    \caption{New medium}
  \end{subfigure}
  \begin{subfigure}{.3\linewidth}
    \includegraphics[width=\columnwidth]{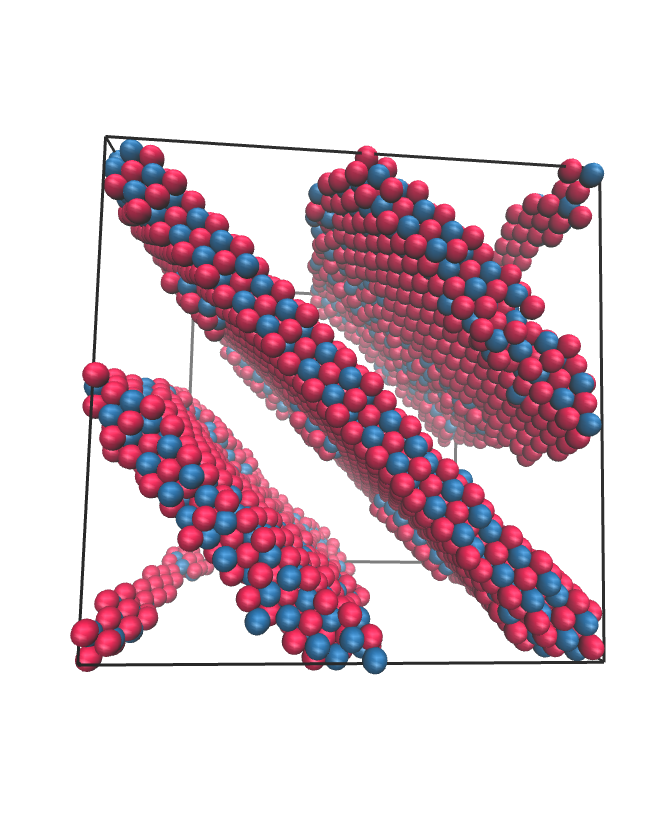}
    \caption{SSP}
  \end{subfigure}
  \caption{Snapshots of configurations for different parametrizations of the nuclear interaction, all with the same thermodynamic conditions: $x = 0.4$, $\rho = 0.05\,\text{fm}^{-3}$ and $T = 0.1\,\text{MeV}$.
    The qualitative differences between CMD medium potential and the other two parametrizations (New Medium and SSP) are evident.
    While the CMD potential shows a \emph{jungle gym} structure, both New Medium and SSP show \emph{lasagna} structures that are slightly different between each other.}
\label{fig:x04_potentials}
\end{figure*}

Different models for the interaction yield different equations of state and, consequently, different configurations.
For comparison, we show in figure~\ref{fig:x01_potentials} different snapshots for the three models we studied: \emph{CMD medium}, \emph{New medium} and \emph{SSP}.
These snapshots are near ground-states, with very low temperature ($T = 0.1\,\text{MeV}$), density $\rho = 0.05\,\text{fm}^{-3}$ and a proton fraction of $x = 0.1$.
The differences are very noticeable: while \emph{CMD medium} potential has no identifiable structure, the \emph{New medium} and \emph{SSP} potentials clearly show agglomerations of proton (due to the binding interaction with neutrons) embedded in a \emph{neutron sea}.
This structure is what we shall call \emph{pre-gnocchi}.
This is the first time such a structure has been identified and it is also a very interesting \emph{qualitative} difference observed among parametrizations of the equation of state.

To compare the potentials in a different configuration, we show in figure~\ref{fig:x04_potentials} different snapshots for the three models we studied: \emph{CMD medium}, \emph{New medium} and \emph{SSP}.
These snapshots are near ground-states, with very low temperature ($T = 0.1\,\text{MeV}$), density $\rho = 0.05\,\text{fm}^{-3}$ and a proton fraction of $x = 0.4$.

\subsection{Asymptotic Mass Distribution}
When the system expands, the structure breaks down into finite fragments.
For long enough times, these fragments remain stable (since they don't interact with each other).
We will refer to this as the asymptotic fragments.

\begin{figure}[h]
  \includegraphics[width=0.8\columnwidth]{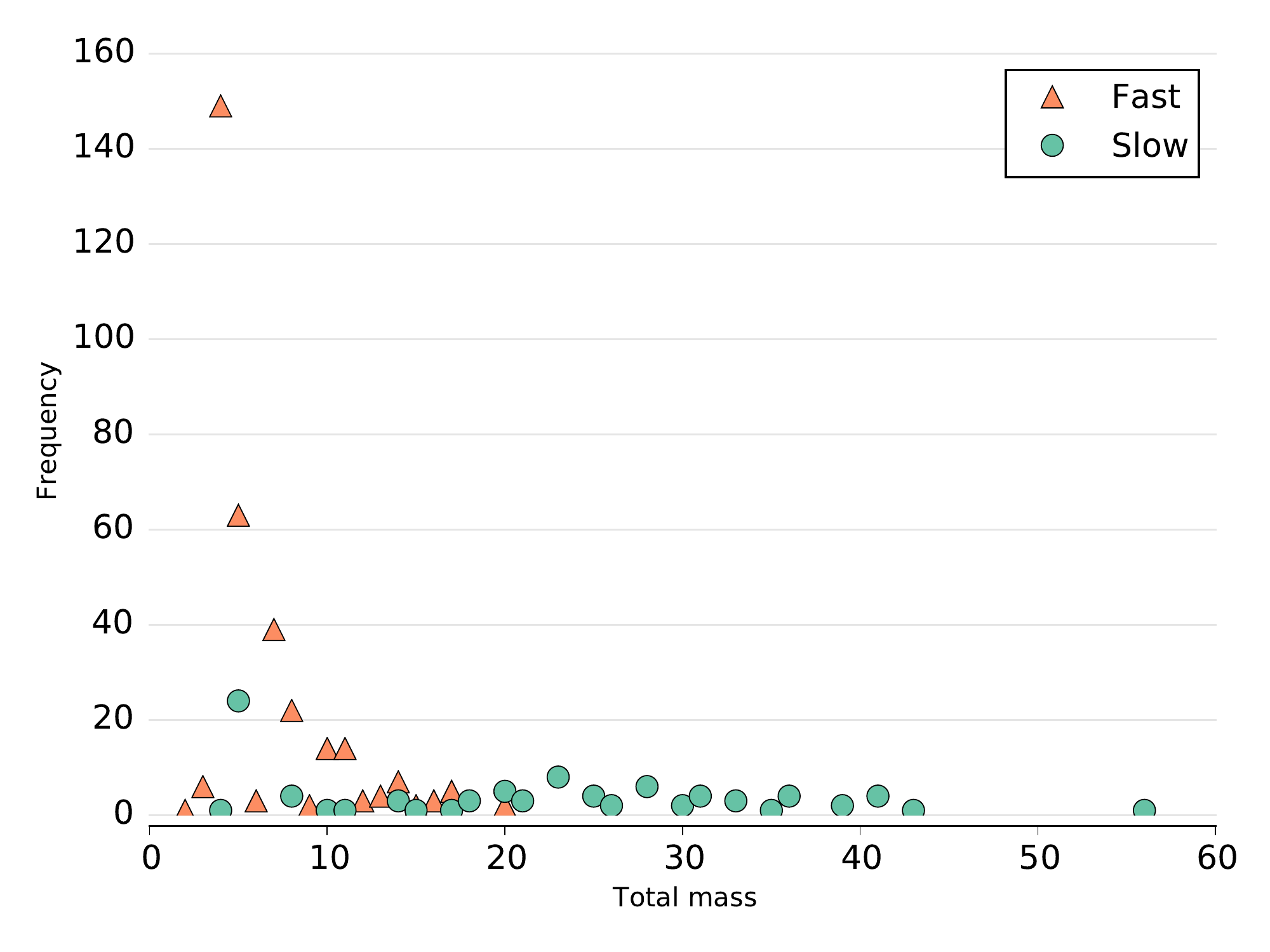}
  \caption{Asymptotic mass distribution for $x = 0.1$, $\rho = 0.05\,\text{fm}^{-3}$ and $T = 0.8\,\text{MeV}$ and two different expansion velocities: \emph{fast} $\dot{\eta} = 0.01\,\text{s}^{-1}$ and \emph{slow} $\dot{\eta} = 0.0001\,\text{s}^{-1}$.}
\label{fig:asymp_preg}
\end{figure}

\begin{figure}[h]
  \includegraphics[width=0.8\columnwidth]{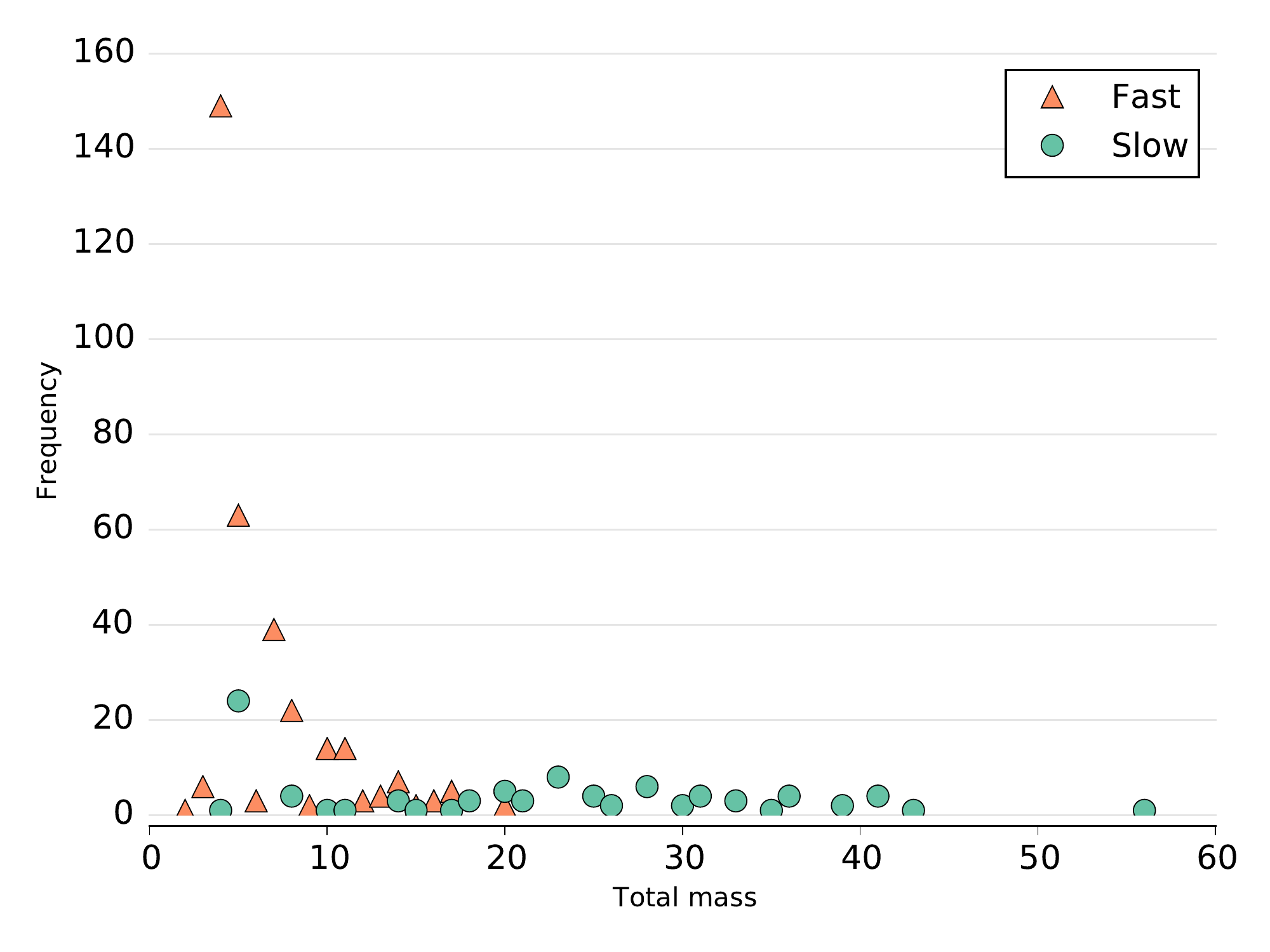}
  \caption{Asymptotic mass distribution for $x = 0.4$, $\rho = 0.05\,\text{fm}^{-3}$ and $T = 0.1\,\text{MeV}$ and two different expansion velocities: \emph{fast} $\dot{\eta} = 0.01\,\text{s}^{-1}$ and \emph{slow} $\dot{\eta} = 0.0001\,\text{s}^{-1}$.}
\label{fig:asymp_las}
\end{figure}

We expanded several initial configurations to find their asymptotic mass distribution.
For the first example, we show in figure~\ref{fig:asymp_preg} the asymptotic mass distribution (calculated with the MSTE algorithm) for $x = 0.1$, $\rho = 0.05\,\text{fm}^{-3}$ and $T = 0.8\,\text{MeV}$ for two expansion velocities: \emph{fast} ($\dot{\eta} = 0.01\,\text{s}^{-1}$) and \emph{slow} ($\dot{\eta} = 0.0001\,\text{s}^{-1}$).
We can see here that the slow expansion allows the existence of fragments with mass of up to 60 (20 of which are protons) while the fast expansion produces smaller fragments of up to 20 (6 protons).
This is an expected behavior, since the faster expansion, the larger the excitation energy.
Therefore, a faster expansion is supposed to break clusters that would otherwise be stable.
A similar behavior can be seen in figure~\ref{fig:asymp_las}, where we expand the system for $x = 0.4$, $\rho = 0.05\,\text{fm}^{-3}$ and $T = 0.1\,\text{MeV}$ for the same fast and slow expansion velocities.
Another relevant characteristic of the asymptotic mass distribution --not shown in the figures due to scale limitations-- is that the fast expansion has a non negligible fraction of lone neutrons (about $4\%$), while the slow expansion hardly presents any ($0.1\%$).

\subsection{Fragment Formation}
\newcommand{\tabfig}[1]{\includegraphics[width=0.25\linewidth]{#1}}

\begin{table*}
  \centering
  \begin{tabular}{cccc}
    \toprule
     & Lasagna (fast expansion) & Lasagna (slow expansion) & Pregnocchi \\
    \midrule
    $\rho = 0.05\,\text{fm}^{-3}$ & \tabfig{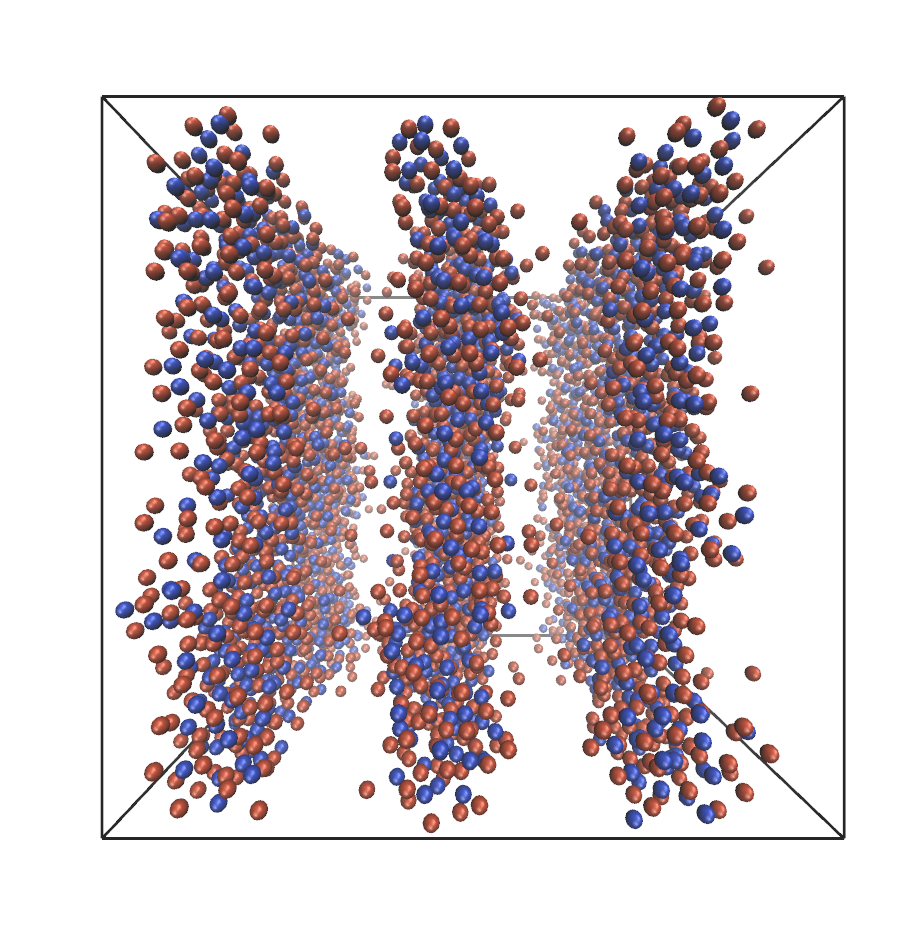} & \tabfig{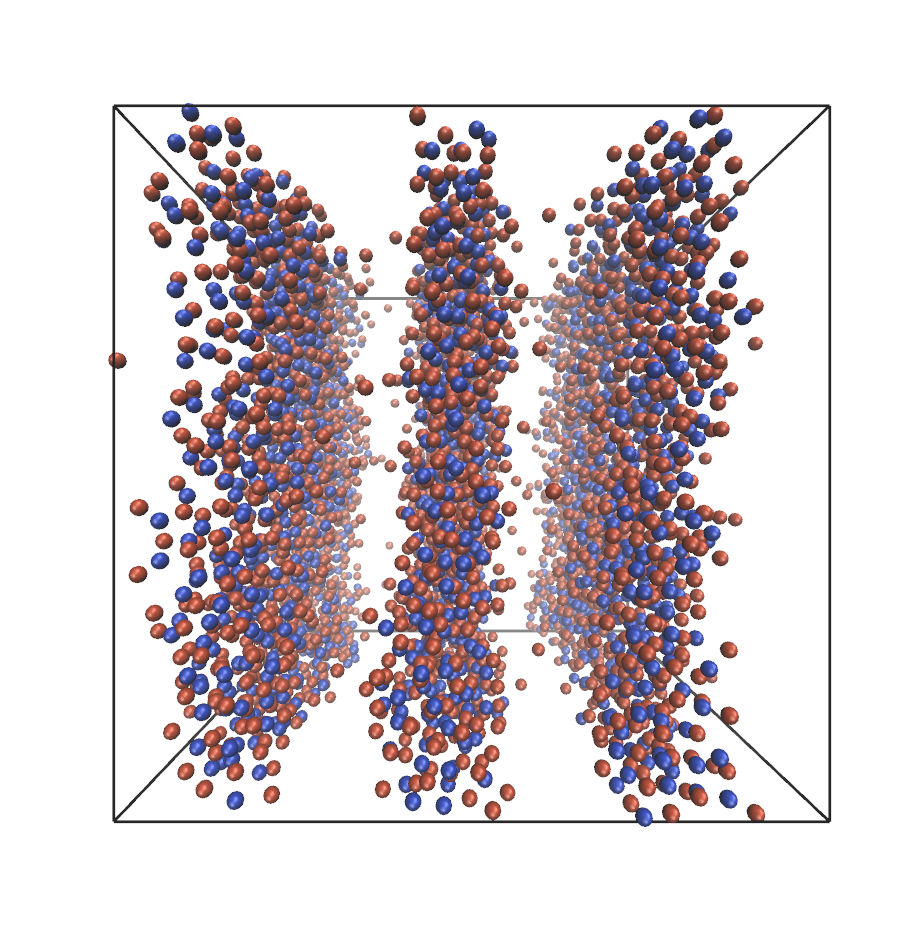} & \tabfig{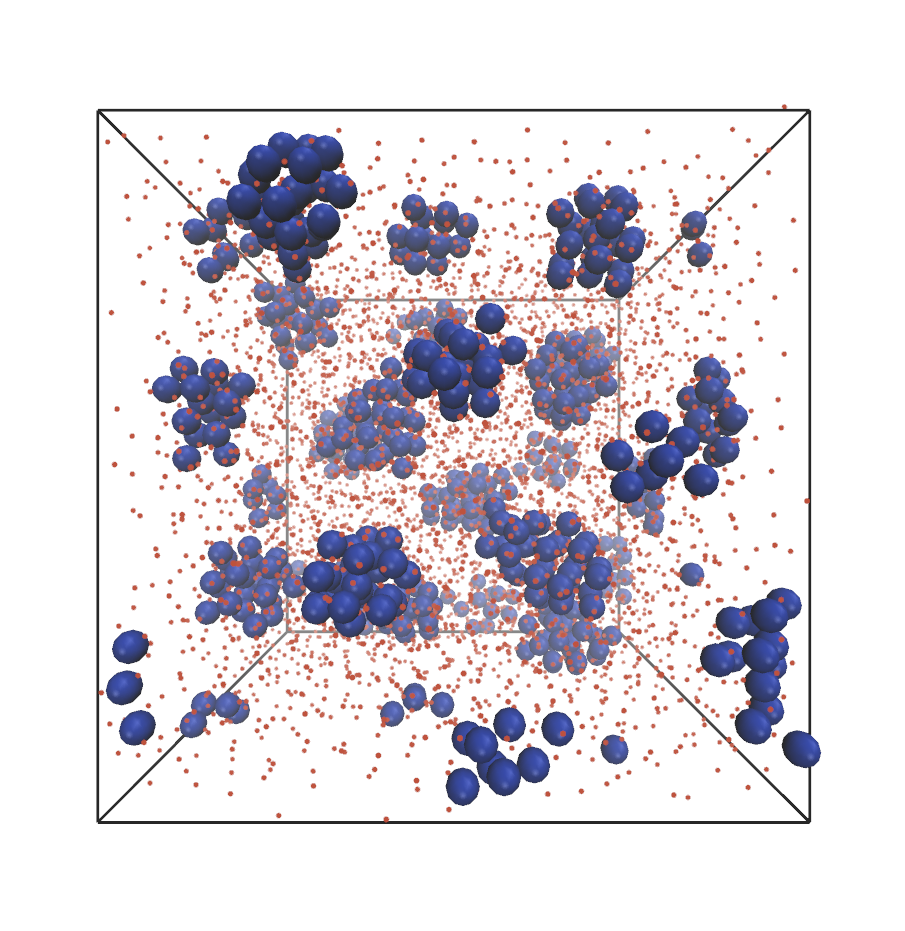} \\
    $\rho = 0.0001\,\text{fm}^{-3}$ & \tabfig{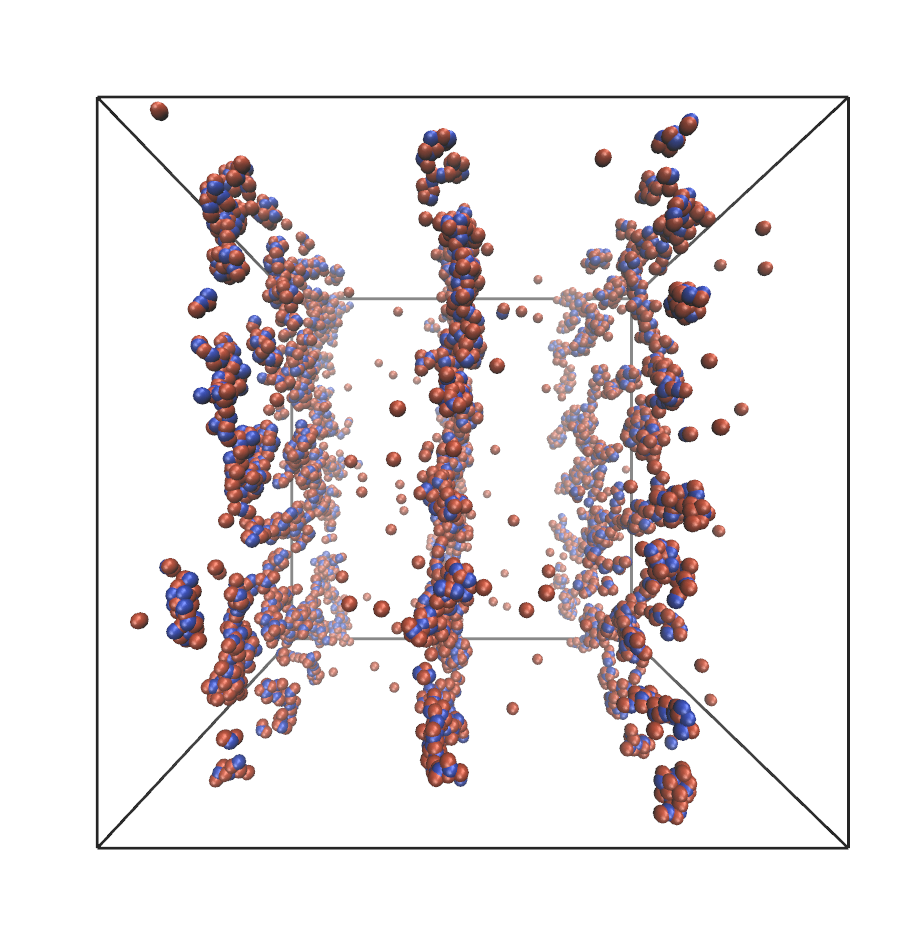} & \tabfig{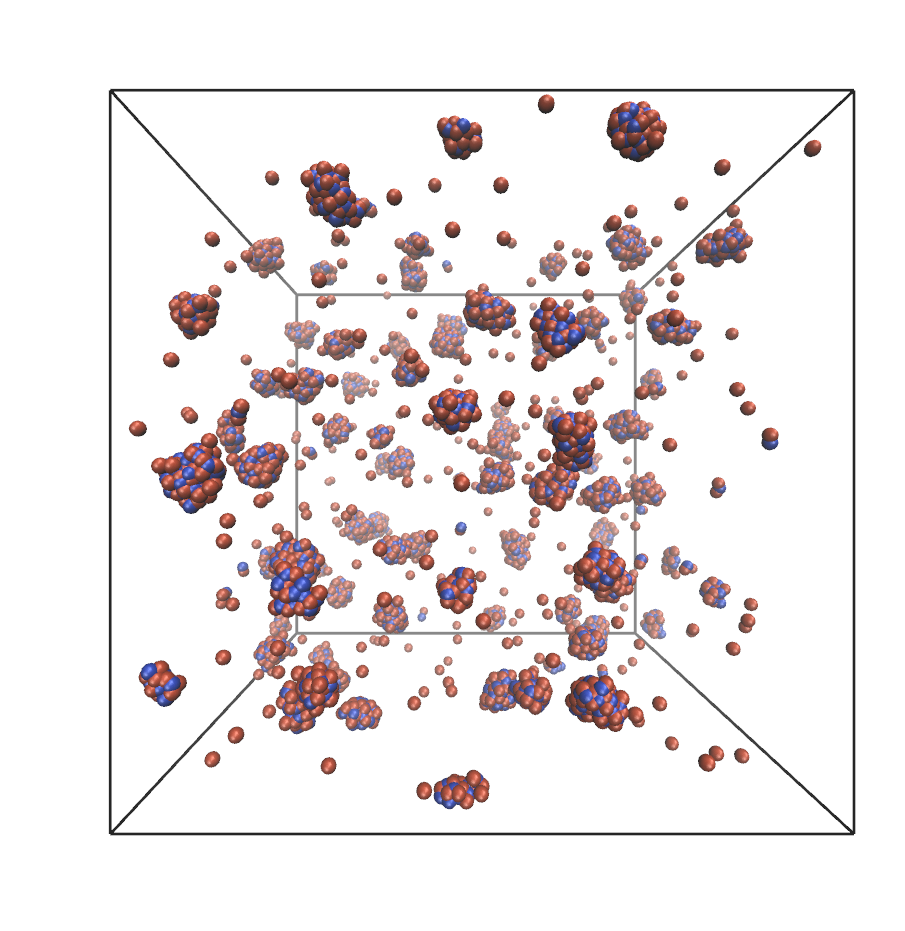} & \tabfig{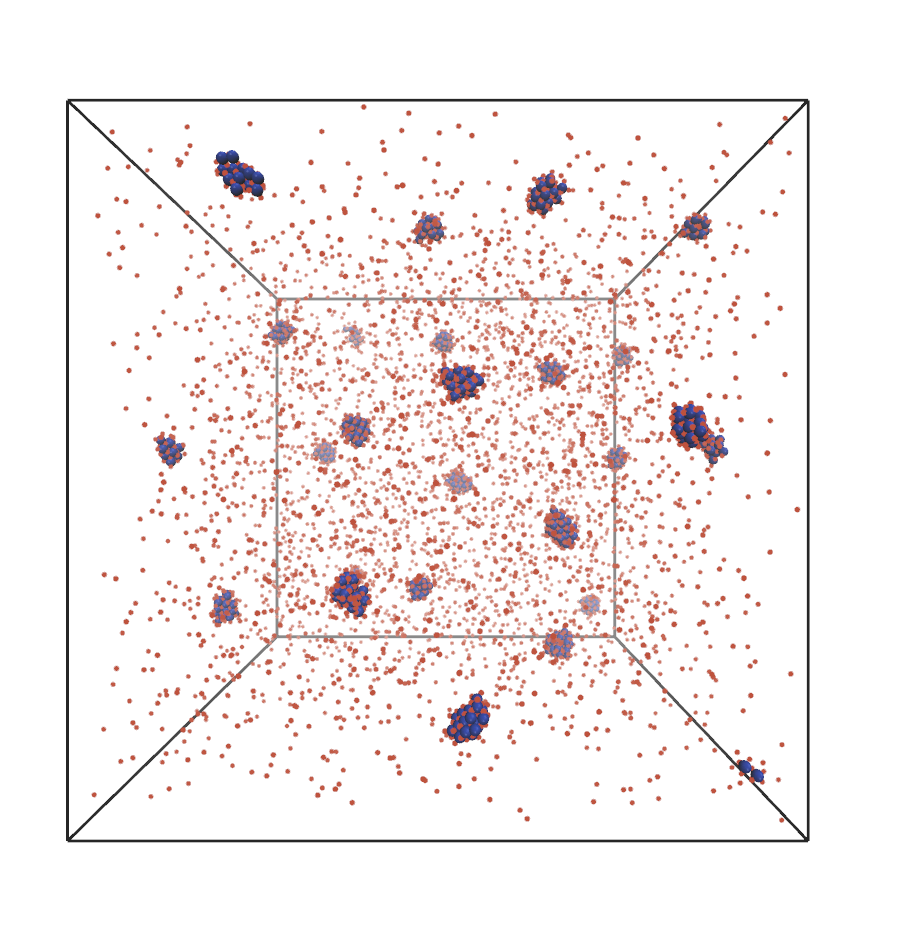} \\
    $\rho = 0.00003\,\text{fm}^{-3}$ & \tabfig{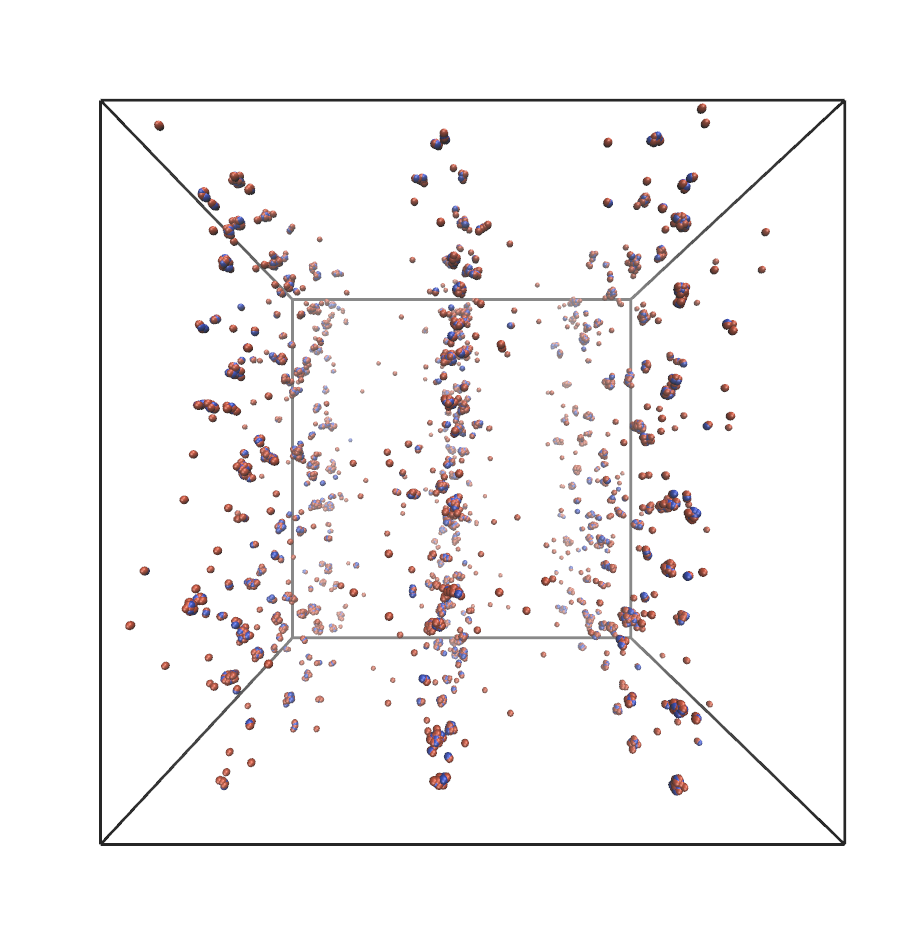} & \tabfig{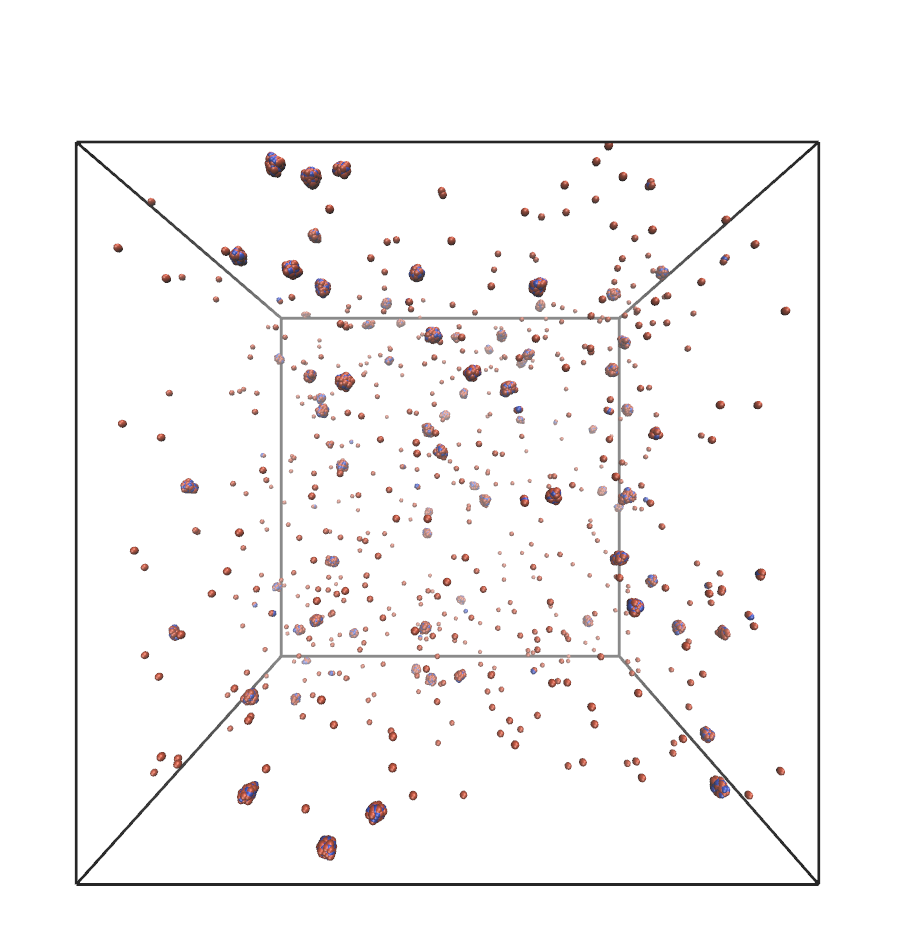} & \tabfig{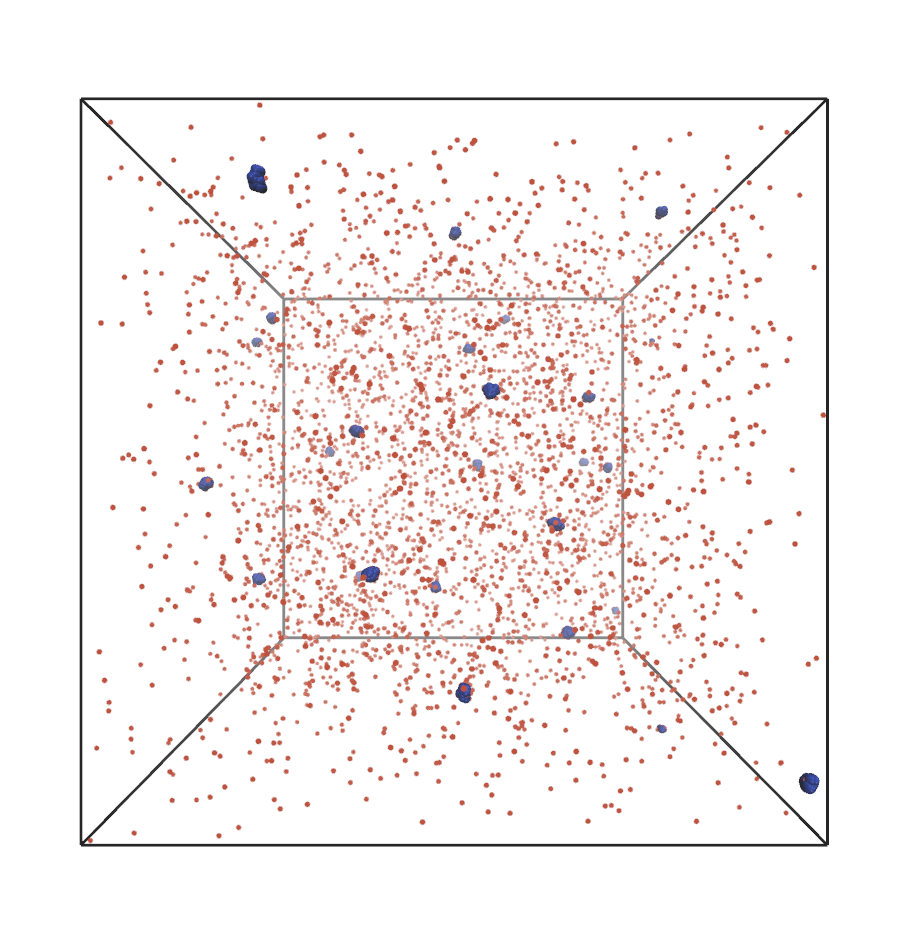} \\
    \bottomrule
  \end{tabular}
  \caption{Three different expansion of Neutron Star Matter: Lasagna (fast expansion): $x = 0.4$, $\dot{\eta} = 0.01\text{s}^{-1}$, $T = 0.8\,\text{MeV}$;
    Lasagna (slow expansion): $x = 0.4$, $\dot{\eta} = 0.0001\text{s}^{-1}$, $T = 0.8\,\text{MeV}$;
    Pregnocchi: $x = 0.1$, $\dot{\eta} = 0.0001\text{s}^{-1}$, $T = 0.1\,\text{MeV}$}
\label{tbl:expansion}
\end{table*}

We now turn to the analysis of some examples of the system evolution in time: when and how are these fragments formed.
We take first the expansion of the system with $x = 0.4$, $\rho = 0.05\,\text{fm}^{-3}$ and $T = 0.5\,\text{MeV}$.
We show in the first two columns of table~\ref{tbl:expansion} the initial and the asymptotic state with the slow and the fast expansion.
While the initial condition is an infinite cluster, in the asymptotic regime we have a fragment distribution with many finite clusters.
It is interesting to note that the fast expansion resembles a mechanical fracture, in which the fragments are formed within each sheet of the lasagna, while the slow expansion looks more like a thermal expansion in which the asymptotic system loses any resemblance to the original structure.
The clusters break into many fragments because their large size cannot withstand the energy associated with the expansion of the system.

A very interesting scenario is the expansion of the system with low proton fraction: $x = 0.1$, $\rho = 0.05\,\text{fm}^{-3}$ and $T = 0.1\,\text{MeV}$.
In the third column of table~\ref{tbl:expansion} we show both the initial condition and the asymptotic configuration for $\dot{\eta} = 0.0001\,\text{s}^{-1}$.

\begin{figure}
  \includegraphics[width=0.8\columnwidth]{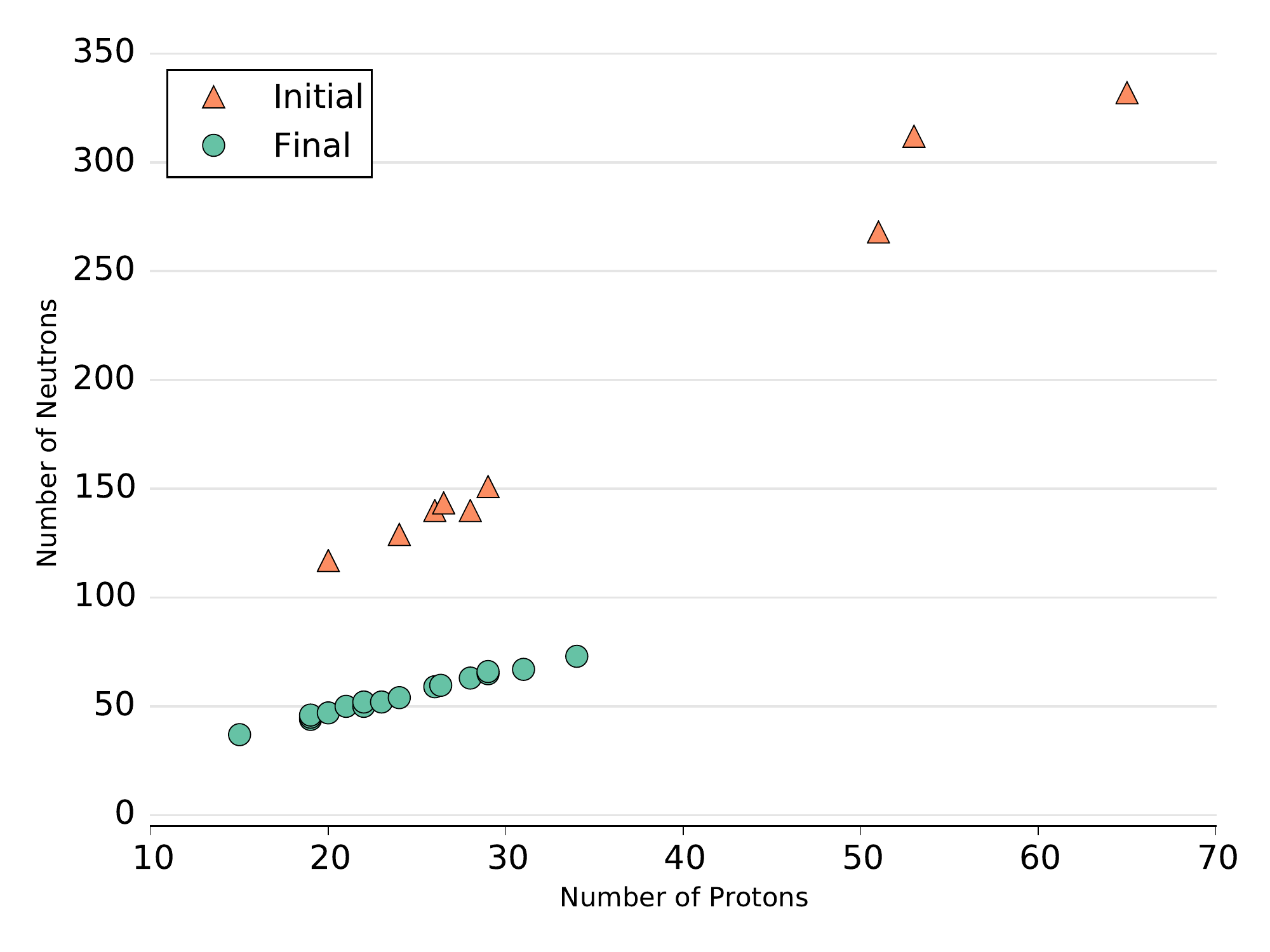}
  \caption{Initial and asymptotic mass distribution for a system with $x = 0.1$, $\rho = 0.05\,\text{fm}^{-3}$ and $T = 0.1\,\text{MeV}$, for a slow expansion ($\dot{\eta} = 0.0001\text{s}^{-1}$), with the MSTE cluster recognition.}
\label{fig:mste_pregnocchi}
\end{figure}

Unlike the previous scenario, there is a clear proton backbone the clusters already exists immersed in a neutron sea.
It can be visually identified when we draw the protons with a much larger size than neutrons, as is in this set of figures.
As the system expands, it is modified.
This raises the question: does the cluster distribution change substantially?
The answer to this question requires a deep analysis of the time evolution of the cluster distribution, and we no longer can rely on a visual inspection; we need to use the cluster recognition algorithms.
Such an analysis has been performed for finite systems for example in Ref.~\cite{dorso_fluctuation_1994, strachan_fragment_1997}.
In figure~\ref{fig:mste_pregnocchi} we show the initial and final configuration with the MSTE algorithm.
Note that the cluster distribution changes radically in both aspects: the size and the proton fraction.
The proton fraction change is to be expected, since as the system expands, less neutrons are within the range of the proton cluster.
However, this effect alone does not explain the change of size: while the initial condition shows a cluster of up to 80 protons, the asymptotic condition's largest cluster is of about 30 protons.
Did a cluster break down while the system was expanding?

\begin{figure}
  \includegraphics[width=0.8\columnwidth]{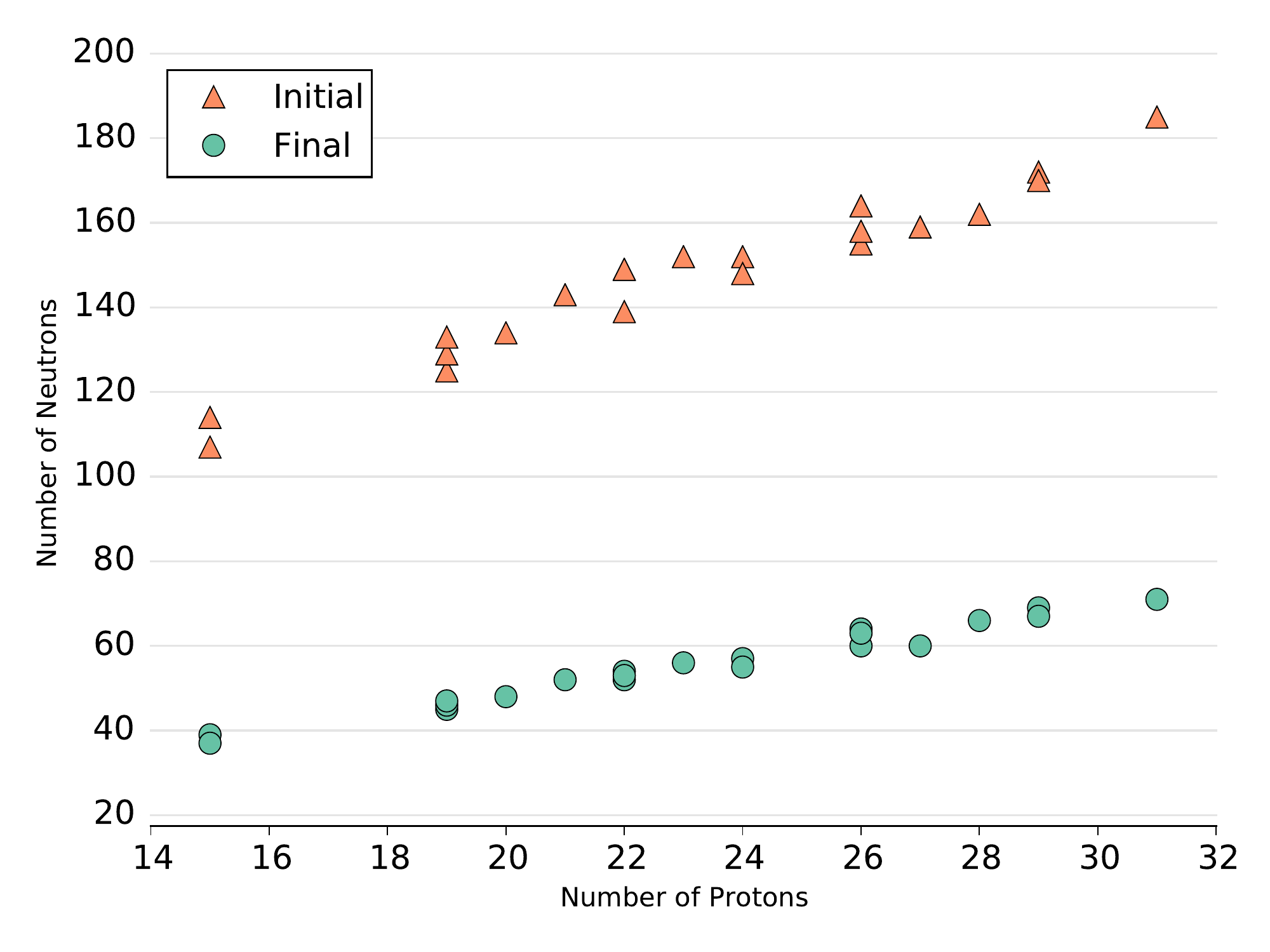}
  \caption{Initial and asymptotic mass distribution for a system with $x = 0.1$, $\rho = 0.05\,\text{fm}^{-3}$ and $T = 0.1\,\text{MeV}$, for a slow expansion ($\dot{\eta} = 0.0001\text{s}^{-1}$), with the proton-MST cluster recognition.}
\label{fig:mst_pregnocchi}
\end{figure}

\begin{figure}
  \includegraphics[width=0.8\columnwidth]{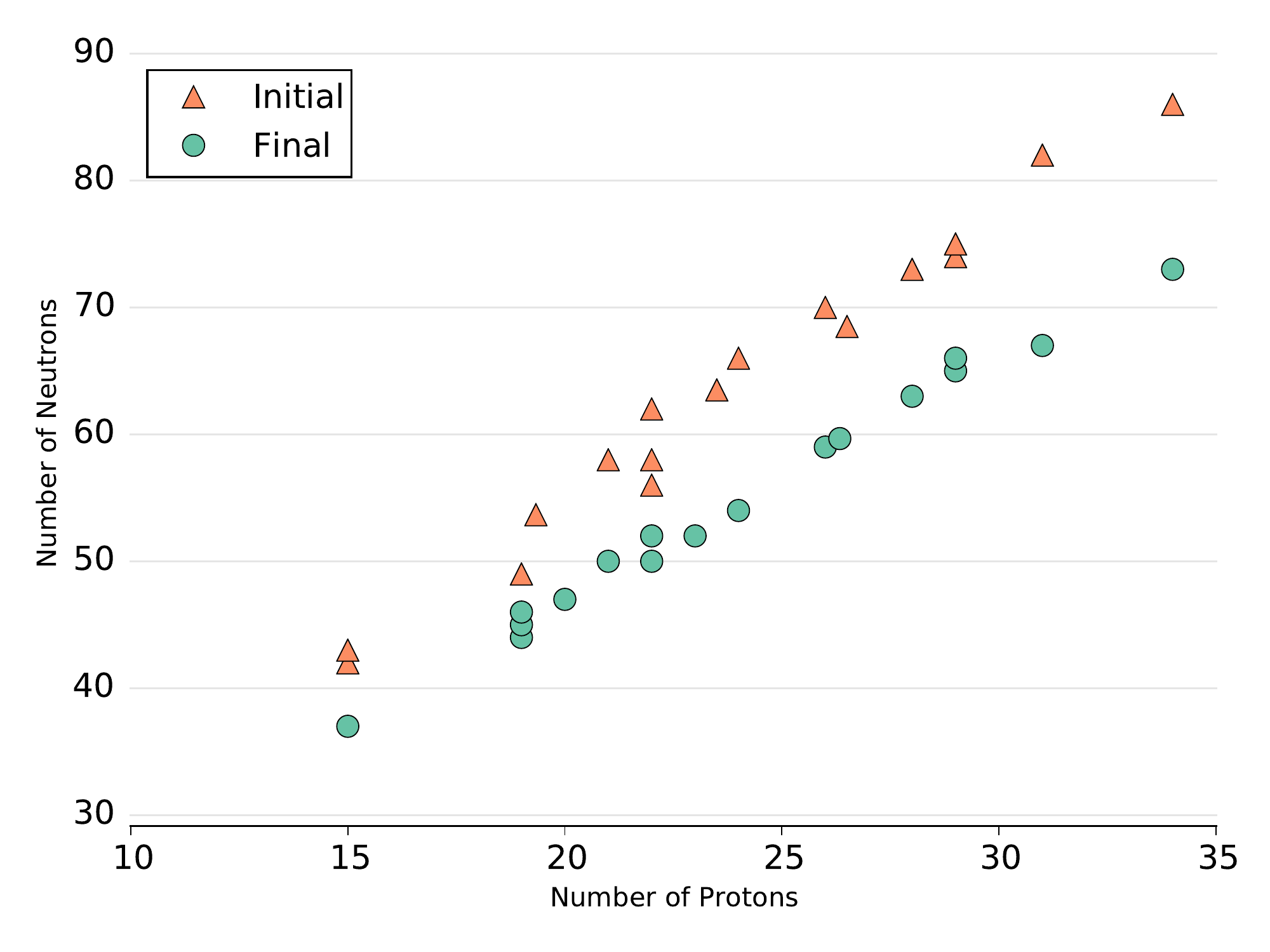}
  \caption{Initial and asymptotic mass distribution for a system with $x = 0.1$, $\rho = 0.05\,\text{fm}^{-3}$ and $T = 0.1\,\text{MeV}$, for a slow expansion ($\dot{\eta} = 0.0001\text{s}^{-1}$), with the ECRA cluster recognition. In comparing with figure~\ref{fig:mst_pregnocchi}, notice the difference in the y-scale.}
\label{fig:ecra_pregnocchi}
\end{figure}

To analyze this, we study the MST distribution of protons alone, shown in figure~\ref{fig:mst_pregnocchi}.
According to this figure, we see that the cluster distribution of protons did not change substantially (only one proton cluster broke down) and effectively the largest cluster has 32 protons.
Does the more theoretically sound ECRA algorithm yield good results?
In figure~\ref{fig:ecra_pregnocchi} we show that actually the ECRA BFM algorithm did yield good results, and identifies the preclusters properly, even finding the proton cluster that broke down.

\begin{figure}[H]  \centering
  \includegraphics[width=0.8\columnwidth]{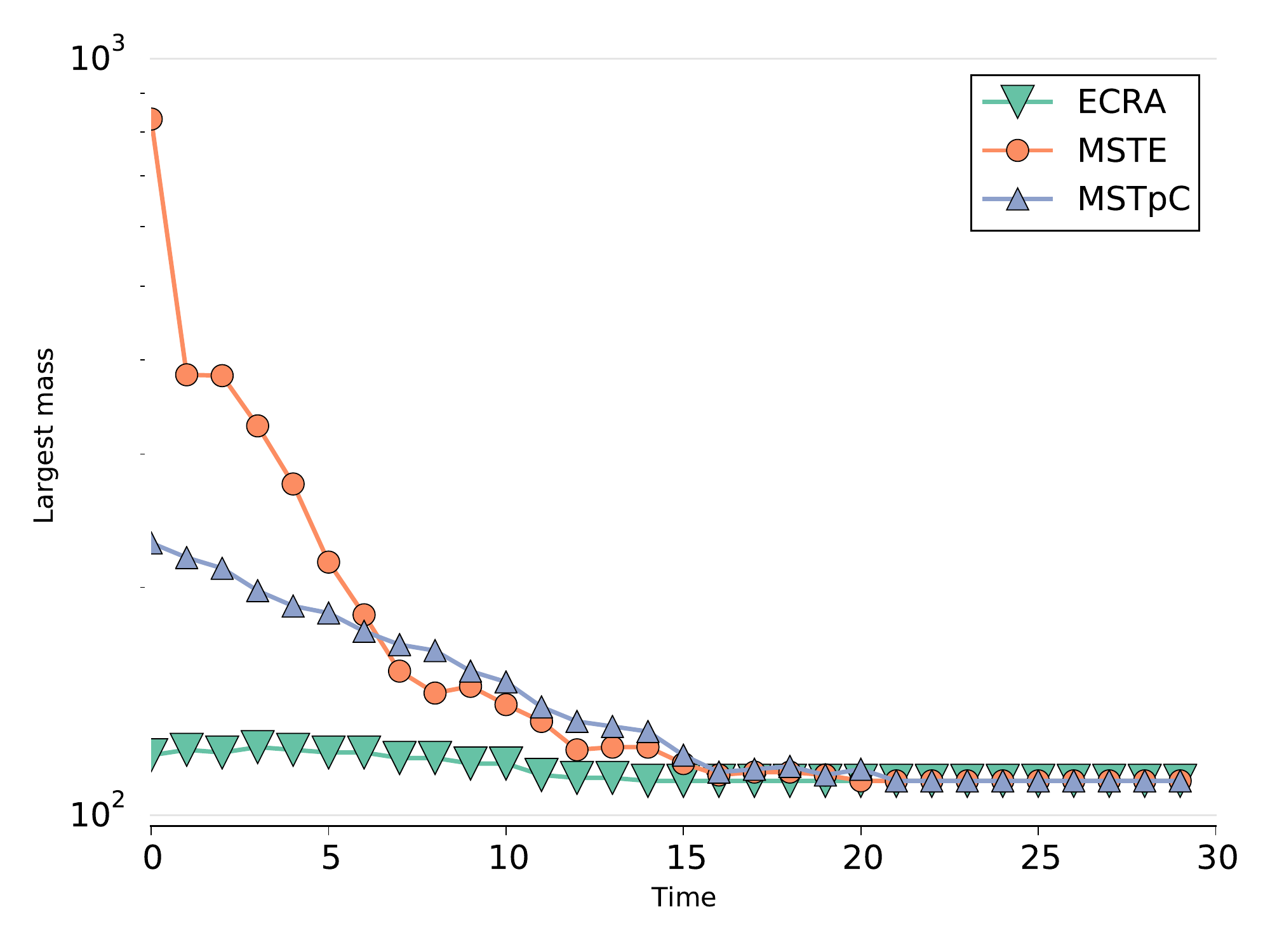}
  \caption{(Color online) Mass of the largest cluster for MSTE, MSTpC and ECRA for the early stages of the evolution, for the \emph{pre-gnocchi} configuration.
    We can see that the ECRA fragment remains relatively stable and stabilizes quickly, while the other two algorithms yield fragments that are always larger and stabilize more slowly.}
\label{fig:lm_pre}
\end{figure}

With these three algorithms in mind, we build three different cluster recognition tools: MSTE, ECRA and MSTpC.
MSTE and ECRA are the regular algorithms, while MSTpC is the proton MST algorithm with the cloud of neutrons that are near each MST cluster.
In figure~\ref{fig:lm_pre} we show the evolution of the size of the largest fragment for the early stages of the evolution for the three clusters: MSTE, MSTpC and ECRA.\@
The figure shows that the ECRA fragment remains relatively stable and stabilizes quickly, while the other two algorithms yield fragments that are always larger and stabilize more slowly.
It is also interesting to note that the MSTpC fragment starts with about 100 neutrons more than the corresponding ECRA fragment, which means that the ECRA fragment is in a very neutron rich environment.
This kind of situation makes the \emph{r-process} more likely to happen.

\begin{figure}[H]  \centering
  \includegraphics[width=0.8\columnwidth]{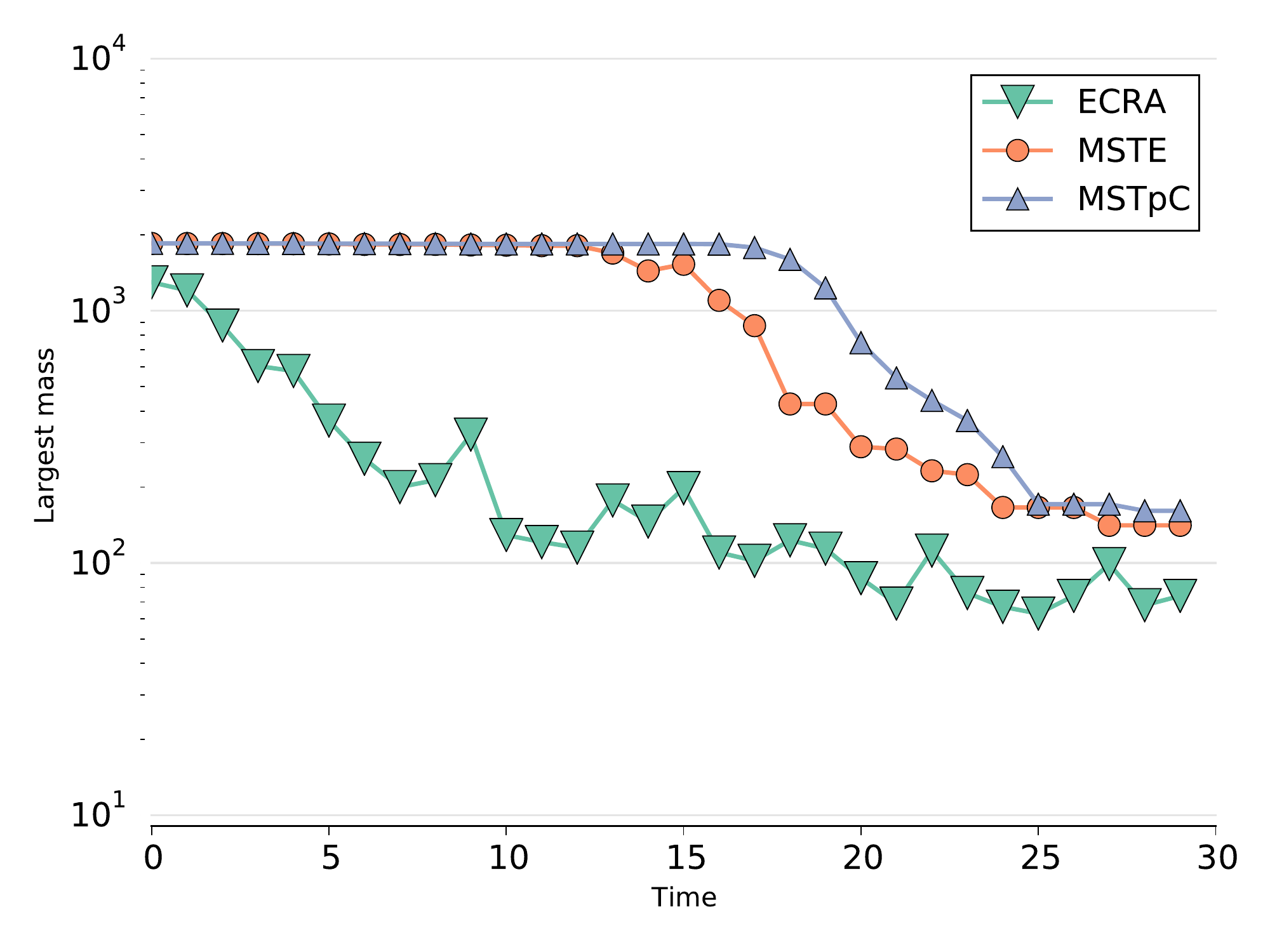}
  \caption{(Color online) Mass of the largest cluster for MSTE, MSTpC and ECRA for the early stages of the evolution, for the \emph{lasagna} configuration.
    See that, as in the previous case, the ECRA algorithm recognizes very early in the expansion the fracture of the clusters.}
\label{fig:lm_las}
\end{figure}

On the other hand, in the expansion of the \emph{lasagna} structure, none of the algorithms for fragment recognition identifies clusters very early on the evolution (see figure~\ref{fig:lm_las}).
At this stage, there is a very large fragment, which is actually infinite.
Nevertheless, the ECRA analysis shows that this fragment breaks down early into many different fragment and, as a result, the mass of the largest fragment decreases drastically with time.
It is interesting to notice that, unlike the \emph{pregnocchi}, in this case the MSTpC algorithm is the one that takes the longest to identify that the infinite cluster breaks down.
This shows us that the ECRA algorithm is also more verstatile to study the early fragment formation.
It is also of interest (not shown) that the proton fraction $x$ of these fragments is relatively stable for the ECRA algorithm, while the other two yield a proton fractio that decreases monotonically with time.

\section{Discussion and Concluding Remarks}\label{sc:conc}

We studied, with molecular dynamics, structural properties of the crust of neutron star through three different potentials.
These potentials involve a nuclear term tailored to reproduce binding energies and compressibilities of nuclear matter plus screened Coulomb interaction.
To analyze the structures formed, we used four different cluster recognition algorithms: MST, MSTE, MSTpC and ECRA-BFM.
With these algorithms we found that of the three potentials, two of them (New Medium and SSP) developed a newly found structure for low proton fractions that we called \emph{pregnocchi}.
This structure consists of proton aggregates formed by the mediation of the attractive $V_{np}$ term of the potential that withstood the expansion.

We also analyzed the expansion of the infinite neutron rich matter described in terms of the little big bang model.
We showed that in general the proper identification of the structure is highly dependent on the algorithm chosen, being ECRA and MSTpC the most suitable to find the structures and ECRA the most stable one.
This approach, combined with different cluster algorithms, allowed us to identify the dynamics of the fragment formation.
The asymptotic state showed a high dependence on the rate of expansion, both in the mass histogram and the spatial distribution of the fragments: for fast enough rates, the expansion was similar to a \emph{mechanical fracture}, where the spatial distribution was heaviliy correlated with the original.
However, for slower rates, the expansion was a \emph{thermal expansion} in which the asymptotic state was relatively homogeneous.
The clusters formed in the slower expansion were much larger than those formed in the fast expansion.
A thorough analysis of the clusters formation dynamics showed that they were formed early in the expansion.
In particular, the novel structure that we have called \emph{pregnocchi} is quite relevant, because according to ECRA analysis these preexistent aggregates evolve in time embedded in a neutron cloud, giving rise to configurations in which \emph{r-proccess} might set in.

\appendix

\section{On the stability of MSTE clusters}
A simple example can be studied to see whether MSTE clusters are always stable.
Consider an interaction
\begin{equation}
V_{ij}(r) =
\begin{cases}
-V_0 &\text{if } r \leq a\\[2ex]
0 &\text{if } r > a.
\end{cases}
\end{equation}

Now we study a set of particles of mass $m$ with positions $r_i = i\,a$ (with $i \in \mathcal{Z}$) so that every particle is at a distance $a$ from its nearest neighbors.
If the velocity is $v_i = i\,v$, each particle will be energetically bound with its neighbors if $v \leq \sqrt{2\,V_0/m}$.
For $2n+1$ particles, with $-n \leq i \leq n$, the kinetic energy of the system will be
\begin{align}
  K_{\text{CM}} &= \sum_{i=-n}^n \frac{1}{2} m\, i^2 v^2\\
  &= \frac{n^3}{3}\,m\,v^2 + \mathcal{O}(n^2)
\end{align}

The potential energy, however, is
\begin{align}
  V_{\text{CM}} &= \sum_{i=-n}^n -i V_0\\
  &= - 2\,n^2\,V_0
\end{align}

It is clear then that for large $n$, no matter the value of $v$, the system will be unstable even though MSTE algorithm recognizes it as a single cluster.
\bibliography{nuclear}

\begin{thebibliography}{49}%
\makeatletter
\providecommand \@ifxundefined [1]{%
 \@ifx{#1\undefined}
}%
\providecommand \@ifnum [1]{%
 \ifnum #1\expandafter \@firstoftwo
 \else \expandafter \@secondoftwo
 \fi
}%
\providecommand \@ifx [1]{%
 \ifx #1\expandafter \@firstoftwo
 \else \expandafter \@secondoftwo
 \fi
}%
\providecommand \natexlab [1]{#1}%
\providecommand \enquote  [1]{``#1''}%
\providecommand \bibnamefont  [1]{#1}%
\providecommand \bibfnamefont [1]{#1}%
\providecommand \citenamefont [1]{#1}%
\providecommand \href@noop [0]{\@secondoftwo}%
\providecommand \href [0]{\begingroup \@sanitize@url \@href}%
\providecommand \@href[1]{\@@startlink{#1}\@@href}%
\providecommand \@@href[1]{\endgroup#1\@@endlink}%
\providecommand \@sanitize@url [0]{\catcode `\\12\catcode `\$12\catcode
  `\&12\catcode `\#12\catcode `\^12\catcode `\_12\catcode `\%12\relax}%
\providecommand \@@startlink[1]{}%
\providecommand \@@endlink[0]{}%
\providecommand \url  [0]{\begingroup\@sanitize@url \@url }%
\providecommand \@url [1]{\endgroup\@href {#1}{\urlprefix }}%
\providecommand \urlprefix  [0]{URL }%
\providecommand \Eprint [0]{\href }%
\providecommand \doibase [0]{http://dx.doi.org/}%
\providecommand \selectlanguage [0]{\@gobble}%
\providecommand \bibinfo  [0]{\@secondoftwo}%
\providecommand \bibfield  [0]{\@secondoftwo}%
\providecommand \translation [1]{[#1]}%
\providecommand \BibitemOpen [0]{}%
\providecommand \bibitemStop [0]{}%
\providecommand \bibitemNoStop [0]{.\EOS\space}%
\providecommand \EOS [0]{\spacefactor3000\relax}%
\providecommand \BibitemShut  [1]{\csname bibitem#1\endcsname}%
\let\auto@bib@innerbib\@empty
%</preamble>
\bibitem [{\citenamefont {Eichler}\ \emph {et~al.}(1989)\citenamefont
  {Eichler}, \citenamefont {Livio}, \citenamefont {Piran},\ and\ \citenamefont
  {Schramm}}]{eichler_nucleosynthesis_1989}%
  \BibitemOpen
  \bibfield  {author} {\bibinfo {author} {\bibfnamefont {D.}~\bibnamefont
  {Eichler}}, \bibinfo {author} {\bibfnamefont {M.}~\bibnamefont {Livio}},
  \bibinfo {author} {\bibfnamefont {T.}~\bibnamefont {Piran}}, \ and\ \bibinfo
  {author} {\bibfnamefont {D.~N.}\ \bibnamefont {Schramm}},\ }\href {\doibase
  10.1038/340126a0} {\bibfield  {journal} {\bibinfo  {journal} {Nature}\
  }\textbf {\bibinfo {volume} {340}},\ \bibinfo {pages} {126} (\bibinfo {year}
  {1989})}\BibitemShut {NoStop}%
\bibitem [{\citenamefont {Freiburghaus}\ \emph {et~al.}(1999)\citenamefont
  {Freiburghaus}, \citenamefont {Rosswog},\ and\ \citenamefont
  {Thielemann}}]{freiburghaus_r-process_1999}%
  \BibitemOpen
  \bibfield  {author} {\bibinfo {author} {\bibfnamefont {C.}~\bibnamefont
  {Freiburghaus}}, \bibinfo {author} {\bibfnamefont {S.}~\bibnamefont
  {Rosswog}}, \ and\ \bibinfo {author} {\bibfnamefont {F.-K.}\ \bibnamefont
  {Thielemann}},\ }\href {\doibase 10.1086/312343} {\bibfield  {journal}
  {\bibinfo  {journal} {ApJ}\ }\textbf {\bibinfo {volume} {525}},\ \bibinfo
  {pages} {L121} (\bibinfo {year} {1999})}\BibitemShut {NoStop}%
\bibitem [{\citenamefont {Tanvir}\ \emph {et~al.}(2013)\citenamefont {Tanvir},
  \citenamefont {Levan}, \citenamefont {Fruchter}, \citenamefont {Hjorth},
  \citenamefont {Hounsell}, \citenamefont {Wiersema},\ and\ \citenamefont
  {Tunnicliffe}}]{tanvir_/`kilonova/_2013}%
  \BibitemOpen
  \bibfield  {author} {\bibinfo {author} {\bibfnamefont {N.~R.}\ \bibnamefont
  {Tanvir}}, \bibinfo {author} {\bibfnamefont {A.~J.}\ \bibnamefont {Levan}},
  \bibinfo {author} {\bibfnamefont {A.~S.}\ \bibnamefont {Fruchter}}, \bibinfo
  {author} {\bibfnamefont {J.}~\bibnamefont {Hjorth}}, \bibinfo {author}
  {\bibfnamefont {R.~A.}\ \bibnamefont {Hounsell}}, \bibinfo {author}
  {\bibfnamefont {K.}~\bibnamefont {Wiersema}}, \ and\ \bibinfo {author}
  {\bibfnamefont {R.~L.}\ \bibnamefont {Tunnicliffe}},\ }\href {\doibase
  10.1038/nature12505} {\bibfield  {journal} {\bibinfo  {journal} {Nature}\
  }\textbf {\bibinfo {volume} {500}},\ \bibinfo {pages} {547} (\bibinfo {year}
  {2013})}\BibitemShut {NoStop}%
\bibitem [{\citenamefont {Lattimer}\ and\ \citenamefont
  {Schramm}(1974)}]{lattimer_black-hole-neutron-star_1974}%
  \BibitemOpen
  \bibfield  {author} {\bibinfo {author} {\bibfnamefont {J.~M.}\ \bibnamefont
  {Lattimer}}\ and\ \bibinfo {author} {\bibfnamefont {D.~N.}\ \bibnamefont
  {Schramm}},\ }\href {\doibase 10.1086/181612} {\bibfield  {journal} {\bibinfo
   {journal} {The Astrophysical Journal Letters}\ }\textbf {\bibinfo {volume}
  {192}},\ \bibinfo {pages} {L145} (\bibinfo {year} {1974})}\BibitemShut
  {NoStop}%
\bibitem [{\citenamefont {Goriely}\ \emph {et~al.}(2011)\citenamefont
  {Goriely}, \citenamefont {Bauswein},\ and\ \citenamefont
  {Janka}}]{goriely_r-process_2011}%
  \BibitemOpen
  \bibfield  {author} {\bibinfo {author} {\bibfnamefont {S.}~\bibnamefont
  {Goriely}}, \bibinfo {author} {\bibfnamefont {A.}~\bibnamefont {Bauswein}}, \
  and\ \bibinfo {author} {\bibfnamefont {H.-T.}\ \bibnamefont {Janka}},\ }\href
  {\doibase 10.1088/2041-8205/738/2/L32} {\bibfield  {journal} {\bibinfo
  {journal} {ApJ}\ }\textbf {\bibinfo {volume} {738}},\ \bibinfo {pages} {L32}
  (\bibinfo {year} {2011})}\BibitemShut {NoStop}%
\bibitem [{\citenamefont {Ravenhall}\ \emph {et~al.}(1983)\citenamefont
  {Ravenhall}, \citenamefont {Pethick},\ and\ \citenamefont
  {Wilson}}]{ravenhall_structure_1983}%
  \BibitemOpen
  \bibfield  {author} {\bibinfo {author} {\bibfnamefont {D.~G.}\ \bibnamefont
  {Ravenhall}}, \bibinfo {author} {\bibfnamefont {C.~J.}\ \bibnamefont
  {Pethick}}, \ and\ \bibinfo {author} {\bibfnamefont {J.~R.}\ \bibnamefont
  {Wilson}},\ }\href {\doibase 10.1103/PhysRevLett.50.2066} {\bibfield
  {journal} {\bibinfo  {journal} {Phys. Rev. Lett.}\ }\textbf {\bibinfo
  {volume} {50}},\ \bibinfo {pages} {2066} (\bibinfo {year}
  {1983})}\BibitemShut {NoStop}%
\bibitem [{\citenamefont {Hashimoto}\ \emph {et~al.}(1984)\citenamefont
  {Hashimoto}, \citenamefont {Seki},\ and\ \citenamefont
  {Yamada}}]{hashimoto_shape_1984}%
  \BibitemOpen
  \bibfield  {author} {\bibinfo {author} {\bibfnamefont {M.-a.}\ \bibnamefont
  {Hashimoto}}, \bibinfo {author} {\bibfnamefont {H.}~\bibnamefont {Seki}}, \
  and\ \bibinfo {author} {\bibfnamefont {M.}~\bibnamefont {Yamada}},\ }\href
  {\doibase 10.1143/PTP.71.320} {\bibfield  {journal} {\bibinfo  {journal}
  {Prog. Theor. Phys.}\ }\textbf {\bibinfo {volume} {71}},\ \bibinfo {pages}
  {320} (\bibinfo {year} {1984})}\BibitemShut {NoStop}%
\bibitem [{\citenamefont {Page}\ \emph {et~al.}(2004)\citenamefont {Page},
  \citenamefont {Lattimer}, \citenamefont {Prakash},\ and\ \citenamefont
  {Steiner}}]{page_minimal_2004}%
  \BibitemOpen
  \bibfield  {author} {\bibinfo {author} {\bibfnamefont {D.}~\bibnamefont
  {Page}}, \bibinfo {author} {\bibfnamefont {J.~M.}\ \bibnamefont {Lattimer}},
  \bibinfo {author} {\bibfnamefont {M.}~\bibnamefont {Prakash}}, \ and\
  \bibinfo {author} {\bibfnamefont {A.~W.}\ \bibnamefont {Steiner}},\ }\href
  {\doibase 10.1086/424844} {\bibfield  {journal} {\bibinfo  {journal} {ApJS}\
  }\textbf {\bibinfo {volume} {155}},\ \bibinfo {pages} {623} (\bibinfo {year}
  {2004})}\BibitemShut {NoStop}%
\bibitem [{\citenamefont {Williams}\ and\ \citenamefont
  {Koonin}(1985)}]{williams_sub-saturation_1985}%
  \BibitemOpen
  \bibfield  {author} {\bibinfo {author} {\bibfnamefont {R.~D.}\ \bibnamefont
  {Williams}}\ and\ \bibinfo {author} {\bibfnamefont {S.~E.}\ \bibnamefont
  {Koonin}},\ }\href {\doibase 10.1016/0375-9474(85)90191-5} {\bibfield
  {journal} {\bibinfo  {journal} {Nuclear Physics A}\ }\textbf {\bibinfo
  {volume} {435}},\ \bibinfo {pages} {844} (\bibinfo {year}
  {1985})}\BibitemShut {NoStop}%
\bibitem [{\citenamefont {Oyamatsu}(1993)}]{oyamatsu_nuclear_1993}%
  \BibitemOpen
  \bibfield  {author} {\bibinfo {author} {\bibfnamefont {K.}~\bibnamefont
  {Oyamatsu}},\ }\href {\doibase 10.1016/0375-9474(93)90020-X} {\bibfield
  {journal} {\bibinfo  {journal} {Nuclear Physics A}\ }\textbf {\bibinfo
  {volume} {561}},\ \bibinfo {pages} {431} (\bibinfo {year}
  {1993})}\BibitemShut {NoStop}%
\bibitem [{\citenamefont {Lorenz}\ \emph {et~al.}(1993)\citenamefont {Lorenz},
  \citenamefont {Ravenhall},\ and\ \citenamefont
  {Pethick}}]{lorenz_neutron_1993}%
  \BibitemOpen
  \bibfield  {author} {\bibinfo {author} {\bibfnamefont {C.~P.}\ \bibnamefont
  {Lorenz}}, \bibinfo {author} {\bibfnamefont {D.~G.}\ \bibnamefont
  {Ravenhall}}, \ and\ \bibinfo {author} {\bibfnamefont {C.~J.}\ \bibnamefont
  {Pethick}},\ }\href {\doibase 10.1103/PhysRevLett.70.379} {\bibfield
  {journal} {\bibinfo  {journal} {Phys. Rev. Lett.}\ }\textbf {\bibinfo
  {volume} {70}},\ \bibinfo {pages} {379} (\bibinfo {year} {1993})}\BibitemShut
  {NoStop}%
\bibitem [{\citenamefont {Cheng}\ \emph {et~al.}(1997)\citenamefont {Cheng},
  \citenamefont {Yao},\ and\ \citenamefont {Dai}}]{cheng_properties_1997}%
  \BibitemOpen
  \bibfield  {author} {\bibinfo {author} {\bibfnamefont {K.~S.}\ \bibnamefont
  {Cheng}}, \bibinfo {author} {\bibfnamefont {C.~C.}\ \bibnamefont {Yao}}, \
  and\ \bibinfo {author} {\bibfnamefont {Z.~G.}\ \bibnamefont {Dai}},\ }\href
  {\doibase 10.1103/PhysRevC.55.2092} {\bibfield  {journal} {\bibinfo
  {journal} {Phys. Rev. C}\ }\textbf {\bibinfo {volume} {55}},\ \bibinfo
  {pages} {2092} (\bibinfo {year} {1997})}\BibitemShut {NoStop}%
\bibitem [{\citenamefont {Watanabe}\ \emph {et~al.}(2000)\citenamefont
  {Watanabe}, \citenamefont {Iida},\ and\ \citenamefont
  {Sato}}]{watanabe_thermodynamic_2000}%
  \BibitemOpen
  \bibfield  {author} {\bibinfo {author} {\bibfnamefont {G.}~\bibnamefont
  {Watanabe}}, \bibinfo {author} {\bibfnamefont {K.}~\bibnamefont {Iida}}, \
  and\ \bibinfo {author} {\bibfnamefont {K.}~\bibnamefont {Sato}},\ }\href
  {\doibase 10.1016/S0375-9474(00)00197-4} {\bibfield  {journal} {\bibinfo
  {journal} {Nuclear Physics A}\ }\textbf {\bibinfo {volume} {676}},\ \bibinfo
  {pages} {455} (\bibinfo {year} {2000})}\BibitemShut {NoStop}%
\bibitem [{\citenamefont {Watanabe}\ and\ \citenamefont
  {Iida}(2003)}]{watanabe_electron_2003}%
  \BibitemOpen
  \bibfield  {author} {\bibinfo {author} {\bibfnamefont {G.}~\bibnamefont
  {Watanabe}}\ and\ \bibinfo {author} {\bibfnamefont {K.}~\bibnamefont
  {Iida}},\ }\href {\doibase 10.1103/PhysRevC.68.045801} {\bibfield  {journal}
  {\bibinfo  {journal} {Phys. Rev. C}\ }\textbf {\bibinfo {volume} {68}},\
  \bibinfo {pages} {045801} (\bibinfo {year} {2003})}\BibitemShut {NoStop}%
\bibitem [{\citenamefont {Nakazato}\ \emph {et~al.}(2009)\citenamefont
  {Nakazato}, \citenamefont {Oyamatsu},\ and\ \citenamefont
  {Yamada}}]{nakazato_gyroid_2009}%
  \BibitemOpen
  \bibfield  {author} {\bibinfo {author} {\bibfnamefont {K.}~\bibnamefont
  {Nakazato}}, \bibinfo {author} {\bibfnamefont {K.}~\bibnamefont {Oyamatsu}},
  \ and\ \bibinfo {author} {\bibfnamefont {S.}~\bibnamefont {Yamada}},\ }\href
  {\doibase 10.1103/PhysRevLett.103.132501} {\bibfield  {journal} {\bibinfo
  {journal} {Phys. Rev. Lett.}\ }\textbf {\bibinfo {volume} {103}},\ \bibinfo
  {pages} {132501} (\bibinfo {year} {2009})}\BibitemShut {NoStop}%
\bibitem [{\citenamefont {Maruyama}\ \emph {et~al.}(1998)\citenamefont
  {Maruyama}, \citenamefont {Niita}, \citenamefont {Oyamatsu}, \citenamefont
  {Maruyama}, \citenamefont {Chiba},\ and\ \citenamefont
  {Iwamoto}}]{maruyama_quantum_1998}%
  \BibitemOpen
  \bibfield  {author} {\bibinfo {author} {\bibfnamefont {T.}~\bibnamefont
  {Maruyama}}, \bibinfo {author} {\bibfnamefont {K.}~\bibnamefont {Niita}},
  \bibinfo {author} {\bibfnamefont {K.}~\bibnamefont {Oyamatsu}}, \bibinfo
  {author} {\bibfnamefont {T.}~\bibnamefont {Maruyama}}, \bibinfo {author}
  {\bibfnamefont {S.}~\bibnamefont {Chiba}}, \ and\ \bibinfo {author}
  {\bibfnamefont {A.}~\bibnamefont {Iwamoto}},\ }\href {\doibase
  10.1103/PhysRevC.57.655} {\bibfield  {journal} {\bibinfo  {journal} {Phys.
  Rev. C}\ }\textbf {\bibinfo {volume} {57}},\ \bibinfo {pages} {655} (\bibinfo
  {year} {1998})}\BibitemShut {NoStop}%
\bibitem [{\citenamefont {Kido}\ \emph {et~al.}(2000)\citenamefont {Kido},
  \citenamefont {Maruyama}, \citenamefont {Niita},\ and\ \citenamefont
  {Chiba}}]{kido_md_2000}%
  \BibitemOpen
  \bibfield  {author} {\bibinfo {author} {\bibfnamefont {T.}~\bibnamefont
  {Kido}}, \bibinfo {author} {\bibfnamefont {T.}~\bibnamefont {Maruyama}},
  \bibinfo {author} {\bibfnamefont {K.}~\bibnamefont {Niita}}, \ and\ \bibinfo
  {author} {\bibfnamefont {S.}~\bibnamefont {Chiba}},\ }\href {\doibase
  10.1016/S0375-9474(99)00736-8} {\bibfield  {journal} {\bibinfo  {journal}
  {Nuclear Physics A}\ }\textbf {\bibinfo {volume} {663–664}},\ \bibinfo
  {pages} {877c} (\bibinfo {year} {2000})}\BibitemShut {NoStop}%
\bibitem [{\citenamefont {Watanabe}\ \emph {et~al.}(2003)\citenamefont
  {Watanabe}, \citenamefont {Sato}, \citenamefont {Yasuoka},\ and\
  \citenamefont {Ebisuzaki}}]{watanabe_structure_2003}%
  \BibitemOpen
  \bibfield  {author} {\bibinfo {author} {\bibfnamefont {G.}~\bibnamefont
  {Watanabe}}, \bibinfo {author} {\bibfnamefont {K.}~\bibnamefont {Sato}},
  \bibinfo {author} {\bibfnamefont {K.}~\bibnamefont {Yasuoka}}, \ and\
  \bibinfo {author} {\bibfnamefont {T.}~\bibnamefont {Ebisuzaki}},\ }\href
  {\doibase 10.1103/PhysRevC.68.035806} {\bibfield  {journal} {\bibinfo
  {journal} {Phys. Rev. C}\ }\textbf {\bibinfo {volume} {68}},\ \bibinfo
  {pages} {035806} (\bibinfo {year} {2003})}\BibitemShut {NoStop}%
\bibitem [{\citenamefont {Horowitz}\ \emph
  {et~al.}(2004{\natexlab{a}})\citenamefont {Horowitz}, \citenamefont
  {Pérez-García}, \citenamefont {Carriere}, \citenamefont {Berry},\ and\
  \citenamefont {Piekarewicz}}]{horowitz_nonuniform_2004}%
  \BibitemOpen
  \bibfield  {author} {\bibinfo {author} {\bibfnamefont {C.}~\bibnamefont
  {Horowitz}}, \bibinfo {author} {\bibfnamefont {M.}~\bibnamefont
  {Pérez-García}}, \bibinfo {author} {\bibfnamefont {J.}~\bibnamefont
  {Carriere}}, \bibinfo {author} {\bibfnamefont {D.}~\bibnamefont {Berry}}, \
  and\ \bibinfo {author} {\bibfnamefont {J.}~\bibnamefont {Piekarewicz}},\
  }\href {\doibase 10.1103/PhysRevC.70.065806} {\bibfield  {journal} {\bibinfo
  {journal} {Phys. Rev. C}\ }\textbf {\bibinfo {volume} {70}},\ \bibinfo
  {pages} {065806} (\bibinfo {year} {2004}{\natexlab{a}})}\BibitemShut
  {NoStop}%
\bibitem [{\citenamefont {Dorso}\ \emph {et~al.}(2012)\citenamefont {Dorso},
  \citenamefont {Giménez~Molinelli},\ and\ \citenamefont
  {López}}]{dorso_topological_2012}%
  \BibitemOpen
  \bibfield  {author} {\bibinfo {author} {\bibfnamefont {C.~O.}\ \bibnamefont
  {Dorso}}, \bibinfo {author} {\bibfnamefont {P.~A.}\ \bibnamefont
  {Giménez~Molinelli}}, \ and\ \bibinfo {author} {\bibfnamefont {J.~A.}\
  \bibnamefont {López}},\ }\href {\doibase 10.1103/PhysRevC.86.055805}
  {\bibfield  {journal} {\bibinfo  {journal} {Phys. Rev. C}\ }\textbf {\bibinfo
  {volume} {86}},\ \bibinfo {pages} {055805} (\bibinfo {year}
  {2012})}\BibitemShut {NoStop}%
\bibitem [{\citenamefont {Alcain}\ \emph
  {et~al.}(2014{\natexlab{a}})\citenamefont {Alcain}, \citenamefont
  {Giménez~Molinelli},\ and\ \citenamefont {Dorso}}]{alcain_beyond_2014}%
  \BibitemOpen
  \bibfield  {author} {\bibinfo {author} {\bibfnamefont {P.~N.}\ \bibnamefont
  {Alcain}}, \bibinfo {author} {\bibfnamefont {P.~A.}\ \bibnamefont
  {Giménez~Molinelli}}, \ and\ \bibinfo {author} {\bibfnamefont {C.~O.}\
  \bibnamefont {Dorso}},\ }\href {\doibase 10.1103/PhysRevC.90.065803}
  {\bibfield  {journal} {\bibinfo  {journal} {Phys. Rev. C}\ }\textbf {\bibinfo
  {volume} {90}},\ \bibinfo {pages} {065803} (\bibinfo {year}
  {2014}{\natexlab{a}})}\BibitemShut {NoStop}%
\bibitem [{\citenamefont {Lenk}\ \emph {et~al.}(1990)\citenamefont {Lenk},
  \citenamefont {Schlagel},\ and\ \citenamefont
  {Pandharipande}}]{lenk_accuracy_1990}%
  \BibitemOpen
  \bibfield  {author} {\bibinfo {author} {\bibfnamefont {R.~J.}\ \bibnamefont
  {Lenk}}, \bibinfo {author} {\bibfnamefont {T.~J.}\ \bibnamefont {Schlagel}},
  \ and\ \bibinfo {author} {\bibfnamefont {V.~R.}\ \bibnamefont
  {Pandharipande}},\ }\href {\doibase 10.1103/PhysRevC.42.372} {\bibfield
  {journal} {\bibinfo  {journal} {Phys. Rev. C}\ }\textbf {\bibinfo {volume}
  {42}},\ \bibinfo {pages} {372} (\bibinfo {year} {1990})}\BibitemShut
  {NoStop}%
\bibitem [{\citenamefont {Dorso}\ \emph {et~al.}(1987)\citenamefont {Dorso},
  \citenamefont {Duarte},\ and\ \citenamefont
  {Randrup}}]{dorso_classical_1987}%
  \BibitemOpen
  \bibfield  {author} {\bibinfo {author} {\bibfnamefont {C.}~\bibnamefont
  {Dorso}}, \bibinfo {author} {\bibfnamefont {S.}~\bibnamefont {Duarte}}, \
  and\ \bibinfo {author} {\bibfnamefont {J.}~\bibnamefont {Randrup}},\ }\href
  {\doibase 10.1016/0370-2693(87)91382-7} {\bibfield  {journal} {\bibinfo
  {journal} {Physics Letters B}\ }\textbf {\bibinfo {volume} {188}},\ \bibinfo
  {pages} {287} (\bibinfo {year} {1987})}\BibitemShut {NoStop}%
\bibitem [{\citenamefont {Dorso}\ and\ \citenamefont
  {Randrup}(1988)}]{dorso_classical_1988}%
  \BibitemOpen
  \bibfield  {author} {\bibinfo {author} {\bibfnamefont {C.}~\bibnamefont
  {Dorso}}\ and\ \bibinfo {author} {\bibfnamefont {J.}~\bibnamefont
  {Randrup}},\ }\href {\doibase 10.1016/0370-2693(88)90030-5} {\bibfield
  {journal} {\bibinfo  {journal} {Physics Letters B}\ }\textbf {\bibinfo
  {volume} {215}},\ \bibinfo {pages} {611} (\bibinfo {year}
  {1988})}\BibitemShut {NoStop}%
\bibitem [{\citenamefont {Hartnack}\ \emph {et~al.}(1989)\citenamefont
  {Hartnack}, \citenamefont {Zhuxia}, \citenamefont {Neise}, \citenamefont
  {Peilert}, \citenamefont {Rosenhauer}, \citenamefont {Sorge}, \citenamefont
  {Aichelin}, \citenamefont {Stöcker},\ and\ \citenamefont
  {Greiner}}]{hartnack_quantum_1989}%
  \BibitemOpen
  \bibfield  {author} {\bibinfo {author} {\bibfnamefont {C.}~\bibnamefont
  {Hartnack}}, \bibinfo {author} {\bibfnamefont {L.}~\bibnamefont {Zhuxia}},
  \bibinfo {author} {\bibfnamefont {L.}~\bibnamefont {Neise}}, \bibinfo
  {author} {\bibfnamefont {G.}~\bibnamefont {Peilert}}, \bibinfo {author}
  {\bibfnamefont {A.}~\bibnamefont {Rosenhauer}}, \bibinfo {author}
  {\bibfnamefont {H.}~\bibnamefont {Sorge}}, \bibinfo {author} {\bibfnamefont
  {J.}~\bibnamefont {Aichelin}}, \bibinfo {author} {\bibfnamefont
  {H.}~\bibnamefont {Stöcker}}, \ and\ \bibinfo {author} {\bibfnamefont
  {W.}~\bibnamefont {Greiner}},\ }\href {\doibase 10.1016/0375-9474(89)90328-X}
  {\bibfield  {journal} {\bibinfo  {journal} {Nuclear Physics A}\ }\textbf
  {\bibinfo {volume} {495}},\ \bibinfo {pages} {303} (\bibinfo {year}
  {1989})}\BibitemShut {NoStop}%
\bibitem [{\citenamefont {Horowitz}\ \emph
  {et~al.}(2004{\natexlab{b}})\citenamefont {Horowitz}, \citenamefont
  {Pérez-García},\ and\ \citenamefont
  {Piekarewicz}}]{horowitz_neutrino-pasta_2004}%
  \BibitemOpen
  \bibfield  {author} {\bibinfo {author} {\bibfnamefont {C.~J.}\ \bibnamefont
  {Horowitz}}, \bibinfo {author} {\bibfnamefont {M.~A.}\ \bibnamefont
  {Pérez-García}}, \ and\ \bibinfo {author} {\bibfnamefont {J.}~\bibnamefont
  {Piekarewicz}},\ }\href {\doibase 10.1103/PhysRevC.69.045804} {\bibfield
  {journal} {\bibinfo  {journal} {Phys. Rev. C}\ }\textbf {\bibinfo {volume}
  {69}},\ \bibinfo {pages} {045804} (\bibinfo {year}
  {2004}{\natexlab{b}})}\BibitemShut {NoStop}%
\bibitem [{\citenamefont {Alcain}\ and\ \citenamefont
  {Dorso}(2017)}]{alcain_neutrino_2017}%
  \BibitemOpen
  \bibfield  {author} {\bibinfo {author} {\bibfnamefont {P.~N.}\ \bibnamefont
  {Alcain}}\ and\ \bibinfo {author} {\bibfnamefont {C.~O.}\ \bibnamefont
  {Dorso}},\ }\href {\doibase 10.1016/j.nuclphysa.2017.02.011} {\bibfield
  {journal} {\bibinfo  {journal} {Nuclear Physics A}\ }\textbf {\bibinfo
  {volume} {961}},\ \bibinfo {pages} {183} (\bibinfo {year}
  {2017})}\BibitemShut {NoStop}%
\bibitem [{\citenamefont {Bonasera}\ \emph {et~al.}(2000)\citenamefont
  {Bonasera}, \citenamefont {Bruno}, \citenamefont {Dorso},\ and\ \citenamefont
  {Mastinu}}]{bonasera_critical_2000}%
  \BibitemOpen
  \bibfield  {author} {\bibinfo {author} {\bibfnamefont {A.}~\bibnamefont
  {Bonasera}}, \bibinfo {author} {\bibfnamefont {M.}~\bibnamefont {Bruno}},
  \bibinfo {author} {\bibfnamefont {C.~O.}\ \bibnamefont {Dorso}}, \ and\
  \bibinfo {author} {\bibfnamefont {F.}~\bibnamefont {Mastinu}},\ }\href
  {http://inis.iaea.org/Search/search.aspx?orig_q=RN:31053937} {\  (\bibinfo
  {year} {2000})}\BibitemShut {NoStop}%
\bibitem [{\citenamefont {Chikazumi}\ \emph {et~al.}(2001)\citenamefont
  {Chikazumi}, \citenamefont {Maruyama}, \citenamefont {Chiba}, \citenamefont
  {Niita},\ and\ \citenamefont {Iwamoto}}]{chikazumi_quantum_2001}%
  \BibitemOpen
  \bibfield  {author} {\bibinfo {author} {\bibfnamefont {S.}~\bibnamefont
  {Chikazumi}}, \bibinfo {author} {\bibfnamefont {T.}~\bibnamefont {Maruyama}},
  \bibinfo {author} {\bibfnamefont {S.}~\bibnamefont {Chiba}}, \bibinfo
  {author} {\bibfnamefont {K.}~\bibnamefont {Niita}}, \ and\ \bibinfo {author}
  {\bibfnamefont {A.}~\bibnamefont {Iwamoto}},\ }\href {\doibase
  10.1103/PhysRevC.63.024602} {\bibfield  {journal} {\bibinfo  {journal} {Phys.
  Rev. C}\ }\textbf {\bibinfo {volume} {63}},\ \bibinfo {pages} {024602}
  (\bibinfo {year} {2001})}\BibitemShut {NoStop}%
\bibitem [{\citenamefont {Caplan}\ \emph {et~al.}(2015)\citenamefont {Caplan},
  \citenamefont {Schneider}, \citenamefont {Horowitz},\ and\ \citenamefont
  {Berry}}]{caplan_pasta_2015}%
  \BibitemOpen
  \bibfield  {author} {\bibinfo {author} {\bibfnamefont {M.~E.}\ \bibnamefont
  {Caplan}}, \bibinfo {author} {\bibfnamefont {A.~S.}\ \bibnamefont
  {Schneider}}, \bibinfo {author} {\bibfnamefont {C.~J.}\ \bibnamefont
  {Horowitz}}, \ and\ \bibinfo {author} {\bibfnamefont {D.~K.}\ \bibnamefont
  {Berry}},\ }\href {\doibase 10.1103/PhysRevC.91.065802} {\bibfield  {journal}
  {\bibinfo  {journal} {Phys. Rev. C}\ }\textbf {\bibinfo {volume} {91}},\
  \bibinfo {pages} {065802} (\bibinfo {year} {2015})}\BibitemShut {NoStop}%
\bibitem [{\citenamefont {Chernomoretz}\ \emph {et~al.}(2002)\citenamefont
  {Chernomoretz}, \citenamefont {Gingras}, \citenamefont {Larochelle},
  \citenamefont {Beaulieu}, \citenamefont {Roy}, \citenamefont {St-Pierre},\
  and\ \citenamefont {Dorso}}]{chernomoretz_quasiclassical_2002}%
  \BibitemOpen
  \bibfield  {author} {\bibinfo {author} {\bibfnamefont {A.}~\bibnamefont
  {Chernomoretz}}, \bibinfo {author} {\bibfnamefont {L.}~\bibnamefont
  {Gingras}}, \bibinfo {author} {\bibfnamefont {Y.}~\bibnamefont {Larochelle}},
  \bibinfo {author} {\bibfnamefont {L.}~\bibnamefont {Beaulieu}}, \bibinfo
  {author} {\bibfnamefont {R.}~\bibnamefont {Roy}}, \bibinfo {author}
  {\bibfnamefont {C.}~\bibnamefont {St-Pierre}}, \ and\ \bibinfo {author}
  {\bibfnamefont {C.~O.}\ \bibnamefont {Dorso}},\ }\href {\doibase
  10.1103/PhysRevC.65.054613} {\bibfield  {journal} {\bibinfo  {journal} {Phys.
  Rev. C}\ }\textbf {\bibinfo {volume} {65}},\ \bibinfo {pages} {054613}
  (\bibinfo {year} {2002})}\BibitemShut {NoStop}%
\bibitem [{\citenamefont {López}\ and\ \citenamefont
  {Dorso}(2000)}]{lopez_lectures_2000}%
  \BibitemOpen
  \bibfield  {author} {\bibinfo {author} {\bibfnamefont {J.~A.}\ \bibnamefont
  {López}}\ and\ \bibinfo {author} {\bibfnamefont {C.}~\bibnamefont {Dorso}},\
  }\href {http://www.worldscientific.com/worldscibooks/10.1142/4169} {\emph
  {\bibinfo {title} {Lectures {Notes} on {Phase} {Transformations} in {Nuclear}
  {Matter}}}}\ (\bibinfo  {publisher} {WORLD SCIENTIFIC},\ \bibinfo {year}
  {2000})\BibitemShut {NoStop}%
\bibitem [{\citenamefont {Barranon}\ \emph {et~al.}(2001)\citenamefont
  {Barranon}, \citenamefont {Dorso},\ and\ \citenamefont
  {Lopez}}]{barranon_searching_2001}%
  \BibitemOpen
  \bibfield  {author} {\bibinfo {author} {\bibfnamefont {A.}~\bibnamefont
  {Barranon}}, \bibinfo {author} {\bibfnamefont {C.~O.}\ \bibnamefont {Dorso}},
  \ and\ \bibinfo {author} {\bibfnamefont {J.~A.}\ \bibnamefont {Lopez}},\
  }\href {http://cat.inist.fr/?aModele=afficheN&cpsidt=13424850} {\bibfield
  {journal} {\bibinfo  {journal} {Revista mexicana de física}\ }\textbf
  {\bibinfo {volume} {47}},\ \bibinfo {pages} {93} (\bibinfo {year}
  {2001})}\BibitemShut {NoStop}%
\bibitem [{\citenamefont {Dorso}\ and\ \citenamefont
  {López}(2001)}]{dorso_selection_2001}%
  \BibitemOpen
  \bibfield  {author} {\bibinfo {author} {\bibfnamefont {C.~O.}\ \bibnamefont
  {Dorso}}\ and\ \bibinfo {author} {\bibfnamefont {J.~A.}\ \bibnamefont
  {López}},\ }\href {\doibase 10.1103/PhysRevC.64.027602} {\bibfield
  {journal} {\bibinfo  {journal} {Phys. Rev. C}\ }\textbf {\bibinfo {volume}
  {64}},\ \bibinfo {pages} {027602} (\bibinfo {year} {2001})}\BibitemShut
  {NoStop}%
\bibitem [{\citenamefont {Barranón}\ \emph {et~al.}(2003)\citenamefont
  {Barranón}, \citenamefont {Cárdenas}, \citenamefont {Dorso},\ and\
  \citenamefont {López}}]{barranon_critical_2003}%
  \BibitemOpen
  \bibfield  {author} {\bibinfo {author} {\bibfnamefont {A.}~\bibnamefont
  {Barranón}}, \bibinfo {author} {\bibfnamefont {R.}~\bibnamefont
  {Cárdenas}}, \bibinfo {author} {\bibfnamefont {C.~O.}\ \bibnamefont
  {Dorso}}, \ and\ \bibinfo {author} {\bibfnamefont {J.~A.}\ \bibnamefont
  {López}},\ }\href {\doibase 10.1556/APH.17.2003.1.8} {\bibfield  {journal}
  {\bibinfo  {journal} {APH N.S., Heavy Ion Physics}\ }\textbf {\bibinfo
  {volume} {17}},\ \bibinfo {pages} {59} (\bibinfo {year} {2003})}\BibitemShut
  {NoStop}%
\bibitem [{\citenamefont {Barrañón}\ \emph {et~al.}(2007)\citenamefont
  {Barrañón}, \citenamefont {Dorso},\ and\ \citenamefont
  {López}}]{barranon_time_2007}%
  \BibitemOpen
  \bibfield  {author} {\bibinfo {author} {\bibfnamefont {A.}~\bibnamefont
  {Barrañón}}, \bibinfo {author} {\bibfnamefont {C.~O.}\ \bibnamefont
  {Dorso}}, \ and\ \bibinfo {author} {\bibfnamefont {J.~A.}\ \bibnamefont
  {López}},\ }\href {\doibase 10.1016/j.nuclphysa.2007.04.008} {\bibfield
  {journal} {\bibinfo  {journal} {Nuclear Physics A}\ }\textbf {\bibinfo
  {volume} {791}},\ \bibinfo {pages} {222} (\bibinfo {year}
  {2007})}\BibitemShut {NoStop}%
\bibitem [{\citenamefont {Barrañón}\ \emph {et~al.}(2004)\citenamefont
  {Barrañón}, \citenamefont {Roa},\ and\ \citenamefont
  {López}}]{barranon_entropy_2004}%
  \BibitemOpen
  \bibfield  {author} {\bibinfo {author} {\bibfnamefont {A.}~\bibnamefont
  {Barrañón}}, \bibinfo {author} {\bibfnamefont {J.}~\bibnamefont {Roa}}, \
  and\ \bibinfo {author} {\bibfnamefont {J.}~\bibnamefont {López}},\ }\href
  {\doibase 10.1103/PhysRevC.69.014601} {\bibfield  {journal} {\bibinfo
  {journal} {Phys. Rev. C}\ }\textbf {\bibinfo {volume} {69}},\ \bibinfo
  {pages} {014601} (\bibinfo {year} {2004})}\BibitemShut {NoStop}%
\bibitem [{\citenamefont {Dorso}\ \emph {et~al.}(2006)\citenamefont {Dorso},
  \citenamefont {Escudero}, \citenamefont {Ison},\ and\ \citenamefont
  {López}}]{dorso_dynamical_2006}%
  \BibitemOpen
  \bibfield  {author} {\bibinfo {author} {\bibfnamefont {C.~O.}\ \bibnamefont
  {Dorso}}, \bibinfo {author} {\bibfnamefont {C.~R.}\ \bibnamefont {Escudero}},
  \bibinfo {author} {\bibfnamefont {M.}~\bibnamefont {Ison}}, \ and\ \bibinfo
  {author} {\bibfnamefont {J.~A.}\ \bibnamefont {López}},\ }\href {\doibase
  10.1103/PhysRevC.73.044601} {\bibfield  {journal} {\bibinfo  {journal} {Phys.
  Rev. C}\ }\textbf {\bibinfo {volume} {73}},\ \bibinfo {pages} {044601}
  (\bibinfo {year} {2006})}\BibitemShut {NoStop}%
\bibitem [{\citenamefont {Dorso}\ \emph {et~al.}(2011)\citenamefont {Dorso},
  \citenamefont {Molinelli},\ and\ \citenamefont
  {López}}]{dorso_isoscaling_2011}%
  \BibitemOpen
  \bibfield  {author} {\bibinfo {author} {\bibfnamefont {C.~A.}\ \bibnamefont
  {Dorso}}, \bibinfo {author} {\bibfnamefont {P.~A.~G.}\ \bibnamefont
  {Molinelli}}, \ and\ \bibinfo {author} {\bibfnamefont {J.~A.}\ \bibnamefont
  {López}},\ }\href {\doibase 10.1088/0954-3899/38/11/115101} {\bibfield
  {journal} {\bibinfo  {journal} {J. Phys. G: Nucl. Part. Phys.}\ }\textbf
  {\bibinfo {volume} {38}},\ \bibinfo {pages} {115101} (\bibinfo {year}
  {2011})}\BibitemShut {NoStop}%
\bibitem [{\citenamefont {Plimpton}(1995)}]{plimpton_fast_1995}%
  \BibitemOpen
  \bibfield  {author} {\bibinfo {author} {\bibfnamefont {S.}~\bibnamefont
  {Plimpton}},\ }\href {\doibase 10.1006/jcph.1995.1039} {\bibfield  {journal}
  {\bibinfo  {journal} {Journal of Computational Physics}\ }\textbf {\bibinfo
  {volume} {117}},\ \bibinfo {pages} {1} (\bibinfo {year} {1995})}\BibitemShut
  {NoStop}%
\bibitem [{\citenamefont {Brown}\ \emph {et~al.}(2012)\citenamefont {Brown},
  \citenamefont {Kohlmeyer}, \citenamefont {Plimpton},\ and\ \citenamefont
  {Tharrington}}]{brown_implementing_2012}%
  \BibitemOpen
  \bibfield  {author} {\bibinfo {author} {\bibfnamefont {W.~M.}\ \bibnamefont
  {Brown}}, \bibinfo {author} {\bibfnamefont {A.}~\bibnamefont {Kohlmeyer}},
  \bibinfo {author} {\bibfnamefont {S.~J.}\ \bibnamefont {Plimpton}}, \ and\
  \bibinfo {author} {\bibfnamefont {A.~N.}\ \bibnamefont {Tharrington}},\
  }\href {\doibase 10.1016/j.cpc.2011.10.012} {\bibfield  {journal} {\bibinfo
  {journal} {Computer Physics Communications}\ }\textbf {\bibinfo {volume}
  {183}},\ \bibinfo {pages} {449} (\bibinfo {year} {2012})}\BibitemShut
  {NoStop}%
\bibitem [{\citenamefont {Fetter}\ and\ \citenamefont
  {Walecka}(2003)}]{fetter_quantum_2003}%
  \BibitemOpen
  \bibfield  {author} {\bibinfo {author} {\bibfnamefont {A.~L.}\ \bibnamefont
  {Fetter}}\ and\ \bibinfo {author} {\bibfnamefont {J.~D.}\ \bibnamefont
  {Walecka}},\ }\href@noop {} {{\selectlanguage {english}\emph {\bibinfo
  {title} {Quantum {Theory} of {Many}-particle {Systems}}}}}\ (\bibinfo
  {publisher} {Courier Dover Publications},\ \bibinfo {year}
  {2003})\BibitemShut {NoStop}%
\bibitem [{\citenamefont {Alcain}\ \emph
  {et~al.}(2014{\natexlab{b}})\citenamefont {Alcain}, \citenamefont
  {Giménez~Molinelli}, \citenamefont {Nichols},\ and\ \citenamefont
  {Dorso}}]{alcain_effect_2014}%
  \BibitemOpen
  \bibfield  {author} {\bibinfo {author} {\bibfnamefont {P.~N.}\ \bibnamefont
  {Alcain}}, \bibinfo {author} {\bibfnamefont {P.~A.}\ \bibnamefont
  {Giménez~Molinelli}}, \bibinfo {author} {\bibfnamefont {J.~I.}\ \bibnamefont
  {Nichols}}, \ and\ \bibinfo {author} {\bibfnamefont {C.~O.}\ \bibnamefont
  {Dorso}},\ }\href {\doibase 10.1103/PhysRevC.89.055801} {\bibfield  {journal}
  {\bibinfo  {journal} {Phys. Rev. C}\ }\textbf {\bibinfo {volume} {89}},\
  \bibinfo {pages} {055801} (\bibinfo {year} {2014}{\natexlab{b}})}\BibitemShut
  {NoStop}%
\bibitem [{\citenamefont {Dorso}\ and\ \citenamefont
  {Randrup}(1993)}]{dorso_early_1993}%
  \BibitemOpen
  \bibfield  {author} {\bibinfo {author} {\bibfnamefont {C.}~\bibnamefont
  {Dorso}}\ and\ \bibinfo {author} {\bibfnamefont {J.}~\bibnamefont
  {Randrup}},\ }\href {\doibase 10.1016/0370-2693(93)91158-J} {\bibfield
  {journal} {\bibinfo  {journal} {Physics Letters B}\ }\textbf {\bibinfo
  {volume} {301}},\ \bibinfo {pages} {328} (\bibinfo {year}
  {1993})}\BibitemShut {NoStop}%
\bibitem [{\citenamefont {Dorso}\ and\ \citenamefont
  {Balonga}(1994)}]{dorso_fluctuation_1994}%
  \BibitemOpen
  \bibfield  {author} {\bibinfo {author} {\bibfnamefont {C.~O.}\ \bibnamefont
  {Dorso}}\ and\ \bibinfo {author} {\bibfnamefont {P.~E.}\ \bibnamefont
  {Balonga}},\ }\href {\doibase 10.1103/PhysRevC.50.991} {\bibfield  {journal}
  {\bibinfo  {journal} {Phys. Rev. C}\ }\textbf {\bibinfo {volume} {50}},\
  \bibinfo {pages} {991} (\bibinfo {year} {1994})}\BibitemShut {NoStop}%
\bibitem [{\citenamefont {Puente}(1999)}]{puente_efficient_1999}%
  \BibitemOpen
  \bibfield  {author} {\bibinfo {author} {\bibfnamefont {A.}~\bibnamefont
  {Puente}},\ }\href {\doibase 10.1016/S0375-9601(99)00534-4} {\bibfield
  {journal} {\bibinfo  {journal} {Physics Letters A}\ }\textbf {\bibinfo
  {volume} {260}},\ \bibinfo {pages} {234} (\bibinfo {year}
  {1999})}\BibitemShut {NoStop}%
\bibitem [{\citenamefont {Alcain}\ and\ \citenamefont
  {Dorso}(2016)}]{alcain_fragmentation_2016}%
  \BibitemOpen
  \bibfield  {author} {\bibinfo {author} {\bibfnamefont {P.~N.}\ \bibnamefont
  {Alcain}}\ and\ \bibinfo {author} {\bibfnamefont {C.~O.}\ \bibnamefont
  {Dorso}},\ }\href {http://arxiv.org/abs/1603.09362} {\bibfield  {journal}
  {\bibinfo  {journal} {arXiv:1603.09362 [nucl-th]}\ } (\bibinfo {year}
  {2016})},\ \bibinfo {note} {arXiv: 1603.09362}\BibitemShut {NoStop}%
\bibitem [{\citenamefont {Dorso}\ and\ \citenamefont
  {Strachan}(1996)}]{dorso_onset_1996}%
  \BibitemOpen
  \bibfield  {author} {\bibinfo {author} {\bibfnamefont {C.~O.}\ \bibnamefont
  {Dorso}}\ and\ \bibinfo {author} {\bibfnamefont {A.}~\bibnamefont
  {Strachan}},\ }\href {\doibase 10.1103/PhysRevB.54.236} {\bibfield  {journal}
  {\bibinfo  {journal} {Phys. Rev. B}\ }\textbf {\bibinfo {volume} {54}},\
  \bibinfo {pages} {236} (\bibinfo {year} {1996})}\BibitemShut {NoStop}%
\bibitem [{\citenamefont {Strachan}\ and\ \citenamefont
  {Dorso}(1997)}]{strachan_fragment_1997}%
  \BibitemOpen
  \bibfield  {author} {\bibinfo {author} {\bibfnamefont {A.}~\bibnamefont
  {Strachan}}\ and\ \bibinfo {author} {\bibfnamefont {C.~O.}\ \bibnamefont
  {Dorso}},\ }\href {\doibase 10.1103/PhysRevC.56.995} {\bibfield  {journal}
  {\bibinfo  {journal} {Phys. Rev. C}\ }\textbf {\bibinfo {volume} {56}},\
  \bibinfo {pages} {995} (\bibinfo {year} {1997})}\BibitemShut {NoStop}%
\end{thebibliography}%
\end{document}